\documentclass[nobibnotes,amsmath,amssymb, aps]{revtex4}

\usepackage{epsfig}
\usepackage{amsfonts}
\usepackage{amsmath}
\usepackage{amssymb}
\usepackage{bm} 
\usepackage[framemethod=PStricks]{mdframed}
\usepackage{lipsum}
\usepackage{float}
\usepackage{color}
\usepackage{pdflscape}
\usepackage{graphicx}
\usepackage{natbib}
\usepackage[counterclockwise]{rotating}

\newcommand{\be}{\begin{equation}}
\newcommand{\ee}{\end{equation}} 
\newcommand{\bea}{\begin{eqnarray}} 
\usepackage{graphicx}
\newcommand{\eea}{\end{eqnarray}}

\DeclareMathOperator{\Tr}{Tr}

\begin{document}

\title{Geometry and Scaling Laws of Excursion and Iso-sets of Enstrophy and Dissipation in Isotropic Turbulence }
\author{Jos\'e Hugo Elsas${}^{1,2}$, Alexander S. Szalay${}^3$ and Charles Meneveau${}^1$}
\affiliation{${}^1$Department of Mechanical Engineering, Johns Hopkins University; 
             ${}^2$Instituto de F\'\i sica, Universidade Federal do Rio de Janeiro;
             ${}^3$Department of Physics and Astronomy and Department of Computer Science, Johns Hopkins University}

\begin{abstract}

  Motivated by interest in the geometry of high intensity events of turbulent flows, we examine spatial correlation functions of 
  sets where turbulent events are particularly intense. These sets are defined using indicator functions on excursion and iso-value
  sets. Their geometric scaling properties  are analyzed by examining possible power-law decay of their radial correlation function.
  We apply the analysis to enstrophy, dissipation, and velocity gradient invariants $Q$ and $R$ and their joint spatial 
  distibutions, using data from a direct numerical simulation of isotropic turbulence at ${\rm Re}_\lambda \approx 430$. 
  While no fractal scaling is found in the inertial range using box-counting in the finite Reynolds number flow considered here, 
  power-law scaling in the inertial range is found in the radial correlation functions. Thus  a geometric 
  characterization in terms of these sets' correlation dimension is possible. Strong dependence on the enstrophy and dissipation threshold 
  is found, consistent with multifractal behavior. Nevertheless the lack of scaling of the box-counting analysis precludes direct quantitative 
  comparisons with earlier work based on the multifractal formalism. Surprising trends, such as a lower correlation dimension for strong dissipation 
  events compared to strong enstrophy events, are observed and interpreted in terms of spatial coherence of vortices in the flow.  
  We show that sets defined by joint conditions on strain and enstrophy, and on $Q$ and $R$,  also display power law scaling of
  correlation functions, providing further characterization of the complex spatial structure of these intersection sets.
\end{abstract}

\maketitle
 
  \section{Introduction} \label{sec:intro}
    Dissipation rate and enstrophy have been observables of great interest in turbulence research due to 
    their dynamical significance for the evolution of the flow and their rich spatial structure and intermittent 
    nature \cite{1995Frisch}. In a view dating back to Kolmogorov \cite{kolmogorov_1941,kolmogorov_1962} and Obukhov \cite{oboukhov_1962}
    the transfer of kinetic energy from large to  small scales proceeds as a self similar cascade process accompanied with 
    increasing intermittency of intense events, and these are often associated with large values of dissipation rate and 
    enstrophy. The presence of power-laws in the velocity spectrum, velocity structure functions and moments of velocity 
    gradients and dissipation are seen as indication of such self-similar behavior.
      
    One of the most common ways to study the resultant intermittent behavior has been through the multifractal 
    formalism. It has its origin in works by Kolmogorov \cite{kolmogorov_1962} and Obukhov \cite{oboukhov_1962} who assumed a 
    lognormal distribution for the dissipation rate, with alternative models proposed by Novikov \& Stewart 
    \cite{NovikovStewart64}, Novikov \cite{1969SPhD...14..104N,1971Novikov,doi:10.1063/1.857629}, 
    Mandelbrot \cite{mandelbrot_1974} and Frisch et al \cite{frisch_sulem_nelkin_1978}. The multifractal approach was 
    formalized explicitly in Benzi et al. \cite{Benzietal1984JPhysA}, making connections to fractal geometry. 
      
    In such methodology, special attention is paid to the power-law scaling of high-order moments of velocity increments 
    (structure functions) or dissipation rates. The approach then invokes a continuos distribution of fractal dimensions 
    $D(h)$ of spatial sets where the velocity increments across a distance $r$ scale with a local Holder exponent $h$, 
    according to $|u(x+r)-u(x)| \sim r^h$ \cite{Benzietal1984JPhysA} ($u(x)$ is a component of the fluid velocity and $r$ is 
    a displacement in the same direction). A  description in terms of local scaling for the  dissipation rate $\epsilon$ and 
    the distribution of its local exponents $\alpha$ with a fractal dimension $f(\alpha)$ has also been used 
    \cite{MeneveauSreenivasanPRL87,MeneveauSreenivasanJFM91}. The multifractal formalism as applied to turbulence has been 
    reviewed  in Refs. \cite{parisiFrisch1985,SreenivasanAnnualReview91,1995Frisch}. In this formalism  the directly measured 
    quantities are the various statistical moments such as $\langle |u(x+r)-u(x)|^p\rangle$  or $\langle \epsilon_r^q\rangle$ 
    (where $\epsilon_r$ is the dissipation averaged in a box of size $r$) while the fractal dimension functions $D(h)$ and 
    $f(\alpha)$ are determined indirectly using the Legendre transformation \cite{1995Frisch} applied to the scaling exponents 
    of the moments. The majority of these prior data analyses were done using one-dimensional experimental surrogates for 
    dissipation rates while only in the last decade have full three-dimensional Direct Numerical Simulations (DNS) begun to 
    approach high enough Reynolds numbers for the possible power law scaling to be discernible \cite{ishihara2009study,Ishihara2008}.
         
    Inspite of the significant success of the multifractal formalism to encapsulate many different phenomena observed since 
    (e.g. multi-point correlations \cite{doi:10.1103/PhysRevE.49.2866}, time correlations \cite{doi:10.1063/1.3623466}, 
    extended self-similarity \cite{doi:0.1103/PhysRevE.48.R29} and varying viscous scales \cite{doi:10.1103/PhysRevE.54.3657}), 
    direct determination of the fractal dimensions as a geometric characterization of the sets of high intensity events has 
    been far less common. An early attempt to study the scale-invariance of histograms of singularities \cite{MENEVEAU1989103} 
    and to deduce the dimension from these scalings have met with mixed success due to strong finite-size corrections and was 
    thus limited to data at very high Reynolds numbers. 
            
    Thus, the status of power-law scaling of geometric features of strong events in turbulence remains unsettled.     
    In the present work we seek a geometric characterization of high-intensity events in turbulence that does not rely on 
    statistical moments of the variable but that identifies the high intensity regions directly based on thresholding of the 
    respective variables of interest. Specifically, we ask whether power-law scaling can be identified for such geometric sets,
    at Reynolds numbers attainable with direct numerical simulations (DNS). In seeking such direct geometric observables and 
    their possible power-law scaling we are also motivated by other fields. For instance, the scaling analysis of geometrical 
    properties of excursion sets has been applied for random sets in probability theory and the theory of random fields 
    \cite{adler2010geometry}, and also have been used for analysis of matter distributions in cosmology 
    \cite{novikov1999minkowski,Mecke:1994ax,calvo2010geometry}. 
    
    In this study we employ a direct way to study regions of varying intensity of enstrophy and dissipation rate: Instead of computing moments of the observable 
    (or box-averaged observable) itself, we first define a geometric set as the set of points where the variables exceed a threshold (or fall into a range of values). 
    We use the indicator function that  takes on a value of 1 inside the set of interest and zero outside. Such geometric sets form convoluted clusters of complicated 
    shapes. For example, high-intensity regions of vorticity are known to be arranged into elongated (worm-like) structures, representing vortices 
    \cite{VincentMeneguzzi91,JimenezetalJFM93}. Nominally each one of these structures would be characterized by a dimension equal to unity. However, 
    it is also well known that these vortices are clustered into regions with possible multi-scaling properties and the scaling of a collection of such vortices is not 
    necessarily obvious. Conversely, high dissipation events are often thought to be distributed along sheets, although again these may have complex spatial structure 
    not necessarily leading to a dimension of two. Once the set is identified, we then compute its two-point correlation function and seek to identify possible inertial-range power-law 
    decay of the tails of the correlation functions. We also perform a direct box-counting analysis of these sets to establish whether direct 
    fractal scaling can be identified in the inertial range of turbulence. 
  
    Furthermore, we extend the analysis to the geometry of sets where both dissipation and enstrophy take on certain values. A 
    ``joint multifractal'' formalism was introduced previously \cite{Meneveauetal90}, but was also based on scaling of joint 
    moments rather than directly based on possible fractal scaling of the geometric objects that arise from joint distributions
    of enstrophy and dissipation. Besides enstrophy and dissipation, we also explore the spatial structure of sets formed by 
    two velocity gradient invariants $Q$ and $R$, observables that have elicited considerable interest in recent years 
    \cite{Meneveau11review}.  The dataset to be considered for this analysis is isotropic turbulence at a Taylor-scale 
    Reynolds number of ${\rm Re}_\lambda \approx 433$ obtained from DNS of forced Navier-Stokes equations \cite{Lietal2008}. 
    We first define the variables of interest and then apply the analysis to the various quantities and joint distributions. 
      
 \section{Definitions and data set} \label{sec:definitions}
           
      The typical observables we are interested in are scalar quantities derived from the velocity gradient tensor $\nabla {\bf u}$. These scalar fields 
      describe the rate of rotation (based on the antisymmetric part of $\nabla {\bf u}$) and the rate of fluid material deformation (via the symmetric part of $\nabla {\bf u}$). 
      Specifically, the two scalar fields that will be considered are defined according to      
      \begin{eqnarray}
        \frac{1}{2} \omega^2({\bf x}) &=& \frac{1}{2} (\nabla \times {\bf u})^2 
                               = \frac{1}{2} \epsilon_{ijk}\epsilon_{klm}\partial_i u_j \partial_l u_m \\
                      S^2({\bf x})    &=& S_{ij} S_{ij} ,~~~ {\rm where} ~~\\
                      S_{ij} &=& \frac{1}{2} (\nabla {\bf u} + \nabla {\bf u}^T)_{ij} = \frac{1}{2} (\partial_i u_j + \partial_j u_i)
      \end{eqnarray}
      
      Note that the dissipation is given by $\epsilon({\bf x}) = 2 \nu S^2({\bf x})$, where $\nu$ is the fluid kinematic 
    viscosity. Hence we refer to $S^2$ as the dissipation henceforth. A significant number of prior studies have focused on 
    dissipation and enstrophy such as \cite{Bershadskii199471,Zhu1996,donzisYeungSree2008,gualaLiberzonTsinoberKinzelbach2007},
    also, other observables of  interest are two scalar invariants of the velocity gradient tensor, called $Q$ and $R$, defined 
    by equations below:  
    
      \begin{eqnarray}\label{eq:QR-def}
                           Q({\bf x})  &=& -\frac{1}{2} \Tr [(\nabla {\bf u})^2] = \frac{1}{2} \omega^2 - S^2 \\
                           R({\bf x})  &=&  - \frac{1}{3} \Tr [(\nabla {\bf u})^3] = - \det \nabla {\bf u}
      \end{eqnarray}
      
     In the flow, the above observables assume a range of values, with $S^2,\omega^2/2$ being non-negative while  
    $Q,R$ can be both positive or negative. Excursion sets are the set of points ${\bf x}$ where such observables are  
    above (or below) a certain threshold, for example $\omega^2/2 > \chi$ ($\chi$ will denote the threshold). We analyze 
    the indicator function $\Theta_\chi({\bf x}) $ of the set of points which satisfy the stated condition. Given a set of 
    interest associated with a threshold $\chi$, we define its indicator function according to 
    
      \begin{equation}\label{eq:indicator-fun}
        \Theta_\chi({\bf x}) =  \begin{cases}
                                                   1, & \text{if } {\bf x} \in  \mbox{set of interest associated with threshold $\chi$}  \\
                                                   0, & \text{otherwise}
                                                 \end{cases}
      \end{equation}
    
    An analogous definition can be made for `interval-based' sets, for example, we can define the region where enstrophy 
    assumes values between $\chi_-$ and $\chi_+$, i.e.  $\chi_- < \omega^2/2 < \chi_+$. From here on, the latter regions will 
    also be referred to as `iso-sets', when $\chi_-$ and $\chi_+$ are very close in value (small bin).
        
      Various statistical features of these sets can be used to characterize their spatial distribution. We are especially 
    interested in the two-point structure of these sets and thus focus on the correlation function of $ \Theta_\chi({\bf x})$  
    defined as:    
      
      \begin{equation}
         {\cal C}_\chi({\bf r}) = \langle \Theta_\chi({\bf x})\Theta_\chi({\bf x}+{\bf r}) \rangle,
      \end{equation}
      where the average is understood as a spatial average over positions ${\bf x}$ when applied to statistically homogeneous flows. 
      Note that we are not subtracting the averages of the indicator 
      function (i.e. we do not define fluctuations of the indicator function but leave it as $0$'s and $1$'s). 
       
      In isotropic turbulence  the more compact  quantity is the angular
      average of the 3D correlation function:
      
      \begin{equation}
        C_\chi(r) = \frac{1}{{\cal C}_\chi({\bf 0})} \int_{S_2} {\cal C}_\chi({\bf r}) ~d\Omega_{\bf r},
      \end{equation}
      i.e.  the normalized, radial correlation function. Phenomenologically, one may expect power-law 
    decaying behavior $C_\chi(r) \sim K_\chi r^{-\gamma_\chi}$ for $r$ in the inertial range due to the expected self-similar 
    behavior of turbulence in that range of scales.  The power law exponent $\gamma_\chi$ is expected, however, to depend  on 
    the threshold. Writing the expected scaling behavior with its dimensional dependencies and a possibly Reynolds number and 
    quantity-dependent prefactor $K_\chi$, we write:
    
      \begin{equation}
        C_\chi(r) \sim K_\chi({\rm Re}_{\lambda})\left(\frac{r}{\eta}\right)^{-\gamma_\chi}; \ \ r >> \eta, \ \ r<L,
      \end{equation}
      where $\eta$ is the Kolmogorov scale and $L$ the integral scale of the flow. The scaling exponent $\gamma_\chi$ is 
    expected to be positive, consistent with a decay of the correlation at increasing distance. A more geometric interpretation 
    of the exponent $\gamma_\chi$ can be invoked by recalling that the correlation dimension  $D$ is defined based on the 
    scaling of the correlation function according to $C_\chi(r) \sim r^{D-E}$ \cite{mandelbrot1982fractal} where $E$ is the 
    dimensionality of the embedding space (here $E=3$). Thus the  dimension corresponding to a correlation decay exponent 
    $\gamma_\chi$ is $D(\chi) = 3-\gamma_\chi$.  
           
      For this work, we chose to perform our analysis on a snapshot from the Johns Hopkins Turbulence Database. The data comes 
    from a DNS of forced isotropic turbulence performance on a $1024^3$ periodic grid, using a pseudo-spetral parallel code. 
    The attained Taylor-scale based Reynolds number is ${\rm Re}_\lambda \approx 433$ time averaged over the database time 
    period, and ${\rm Re}_\lambda \approx 426$ for the specific timestep  used in the present analysis ($t=0.0$). The domain 
    is a periodic cube of size $[0,2\pi]^3$, in which the data frames were stored after the simulation reached a statistically 
    stationary state. Additional details of the dataset can be found in \cite{Lietal2008,Johnsonetal16}. In order to establish 
    the scaling range corresponding to the turbulence inertial range for comparison with present results we evaluate the 
    longitudinal structure function as an average over the three Cartesian directions:
    
      \begin{equation}
        D_{\rm LL}(r) = \frac{1}{ u^2_{rms}} \frac{1}{3} \sum \limits_{i=0}^3 \langle \left( u_i({\bf x}+r{\bf e}_i)-u_i({\bf x})\right)^2 \rangle,
      \end{equation} 
    where ${\bf e}_i$ is the unit vector in the direction of the velocity component $u_i$, and 
    $u^2_{rms} = \frac{1}{3}\sum_{i=0}^3 u_i^2$ is the square of the RMS velocity. 
  
      In order to evaluate enstrophy and dissipation, the velocity gradients  are calculated  with spectral accuracy using 
    Fast Fourier Transform (FFT). For our present analysis we did not use the databases' finite differencing or Spline 
    differencing tools since these are less accurate compared to spectral methods that were also used during the DNS. As 
    further explained in  Appendix A,  the analysis was done on a server near the database using notebooks provided by a 
    dedicated compute environment (the SciServer system).  For differentiation,  a 3D FFT operation is applied to the velocity 
    field to obtain the velocity field in Fourier space, then the components are multiplied by the respective wavenumbers 
    ($i k_j$) to obtain the velocity gradient in the $x_j$ direction. Finally the inverse FFT is applied to obtain the velocity 
    gradients in physical space. This allows us to obtain $A_{ij} = \partial_j u_i$ data from the velocity field ${\bf u}$. 
    The observables we are interested in (dissipation $S^2$, enstrophy $\omega^2/2$,  $Q$ and $R$) are then computed in 
    physical space.
    
      To compute the correlation functions efficiently in 3D,  a 3D FFT is applied to $\Theta_\chi({\bf x})$ over the $1024^3$ 
    data-cube, yielding $\tilde \Theta_\chi({\bf k})$ in Fourier space. To $\tilde \Theta_\chi \tilde \Theta_\chi^*$ is then 
    applied the inverse Fourier transform, resulting in ${\cal C}({\bf r})$, which is the full 3D two-point correlation 
    function. The radial integration is done by evaluating a histogram based on the radial $\sqrt{|{\bf r}|^2}$ values 
    computed over the resulting grid. This effectively performs the  angular average by dividing the weighted average of the 
    two-point correlation and the base $\sqrt{|{\bf r}|^2}$ histogram. More details are provided in Appendix A.
    
   \section{Excursion set analysis} \label{sec:excursion}
\begin{figure}[H]
           \begin{center}
       \begin{minipage}{0.4\linewidth}
         \includegraphics[width=\linewidth]{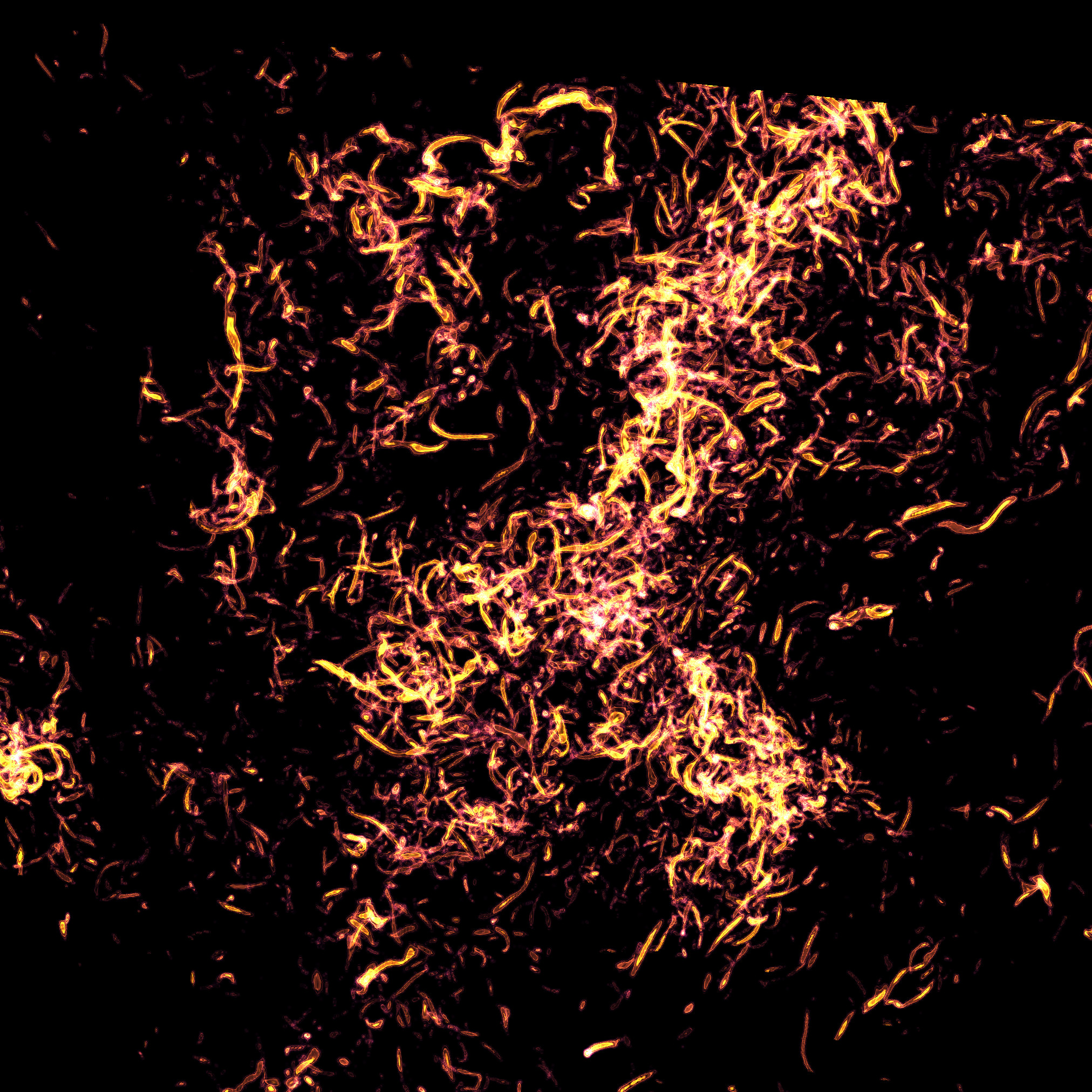}            
       \end{minipage}
       \begin{minipage}{0.4\linewidth}
         \includegraphics[width=\linewidth]{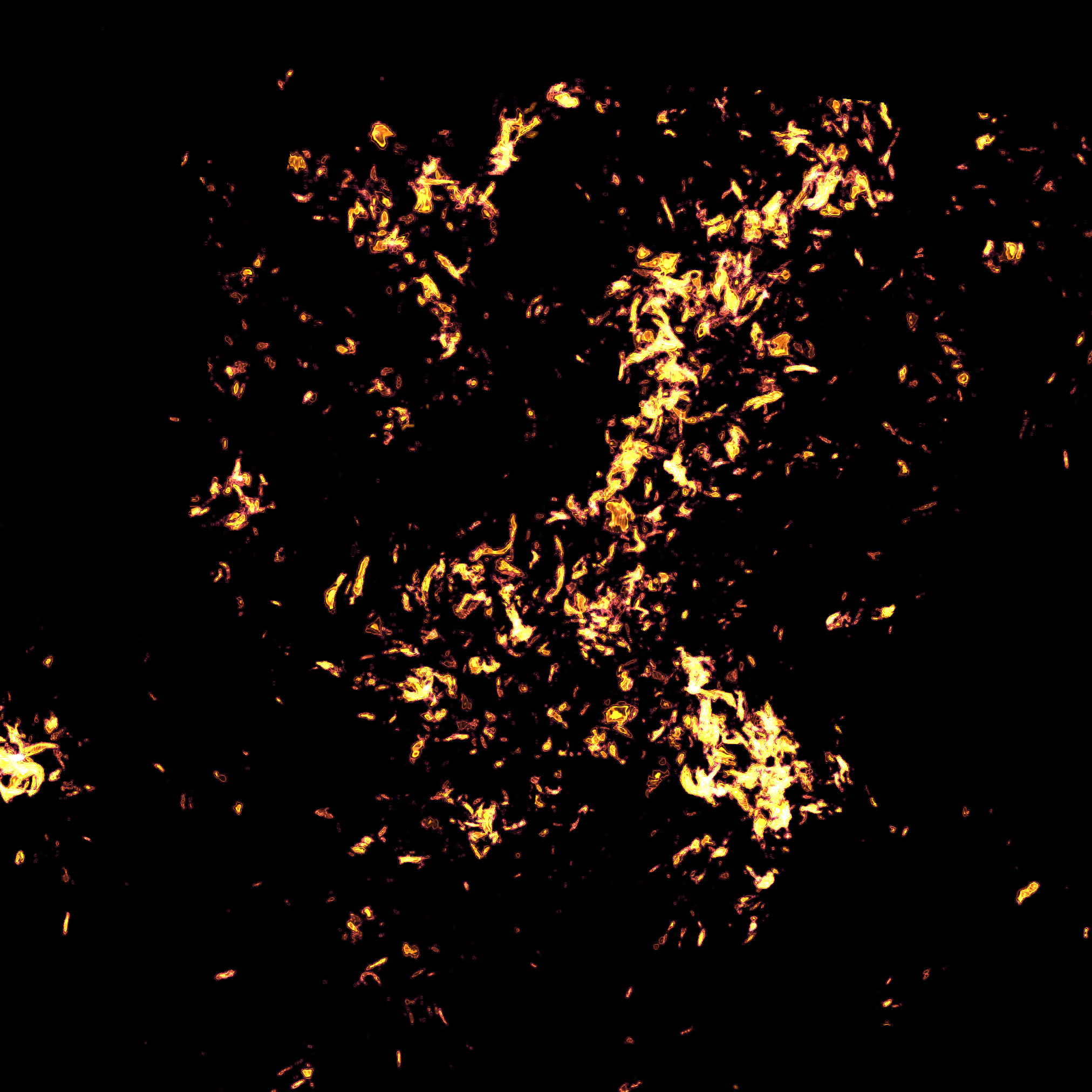}            
       \end{minipage}
             \end{center}
       \caption{Volume rendering of (a) the enstrophy excursion set corresponding to the function \\
                $\frac{1}{2}\omega^2_{ex}({\bf x}) = \frac{1}{2} \omega^2({\bf x}) \Theta_{20\langle S^2\rangle}^\omega({\bf x})$ 
                and (b) of dissipation excursion set corresponding to  
                the function $S^2_{ex}({\bf x}) = S^2({\bf x}) \Theta_{20\langle S^2\rangle}^S({\bf x})$
                on a $512^3$ subset of the full dataset, with $1/8$ in volume. The visualizations were generated using the YT-project 
                python visualization library \cite{yt-project}. }
        \label{fig:excursion-20-vol}
      \end{figure}

      The excursion set indicator function for a given scalar field $A$, like enstrophy $A = \omega^2/2$ or dissipation 
      $A = S^2$, is defined as:
      
      \begin{equation}\label{eq:indicator-fun-1}
        \Theta^{A}_\chi({\bf x}) = H(A({\bf x})-\chi) = \begin{cases}
                                                   1, & \text{if } A({\bf x})\geq \chi \\
                                                   0, & \text{otherwise},
                                                 \end{cases}
      \end{equation}
      
      where $\chi$ is the threshold applied on the scalar $A$. We begin by considering enstrophy excursion sets. Figure 
    \ref{fig:excursion-20-vol} shows a volume rendering of the scalar function corresponding to enstrophy above the threshold 
    $\chi = 20\langle S^2\rangle$, given by 
    $\frac{1}{2}\omega^2_{ex}({\bf x}) = \frac{1}{2}\omega^2({\bf x}) \Theta_{20\langle S^2\rangle}({\bf x})$. 
    Note for consistency all threshold values are indicated as multiples of $\langle S^2\rangle$ which is equally relevant to 
    enstrophy here since in   isotropic turbulence  $\langle S^2\rangle =   \frac{1}{2}\langle\omega^2\rangle$. As can be seen 
    in Fig.  \ref{fig:excursion-20-vol}(a), this set has a very rich structure with familiar elongated strong vortices visible.  
        
      As a comparison, we also present in figure \ref{fig:excursion-20-vol}(b) the visualization of the dissipation field for 
    the same threshold, i.e. $S^2_{ex}({\bf x}) = S^2({\bf x}) \Theta_{20\langle S^2\rangle}^S({\bf x})$, which provides us 
    with some useful insights. The most striking feature is that the overall geometric distribution of high intensity 
    regions for dissipation closely follows the ones for high enstrophy, though the smaller scale details differ. The second 
    feature is that at small-scales dissipation appears to be less 1-D like and more sheet like, but that when viewed
    at larger scales, in comparison to its enstrophy counterpart. As will be seen, this fact will be visible also 
    quantitatively in the correlation function results. 
     
  \subsection{Correlation function based scaling}
  
\begin{figure}[H]
        \begin{minipage}{0.49\linewidth}
            \includegraphics[width=0.92\linewidth]{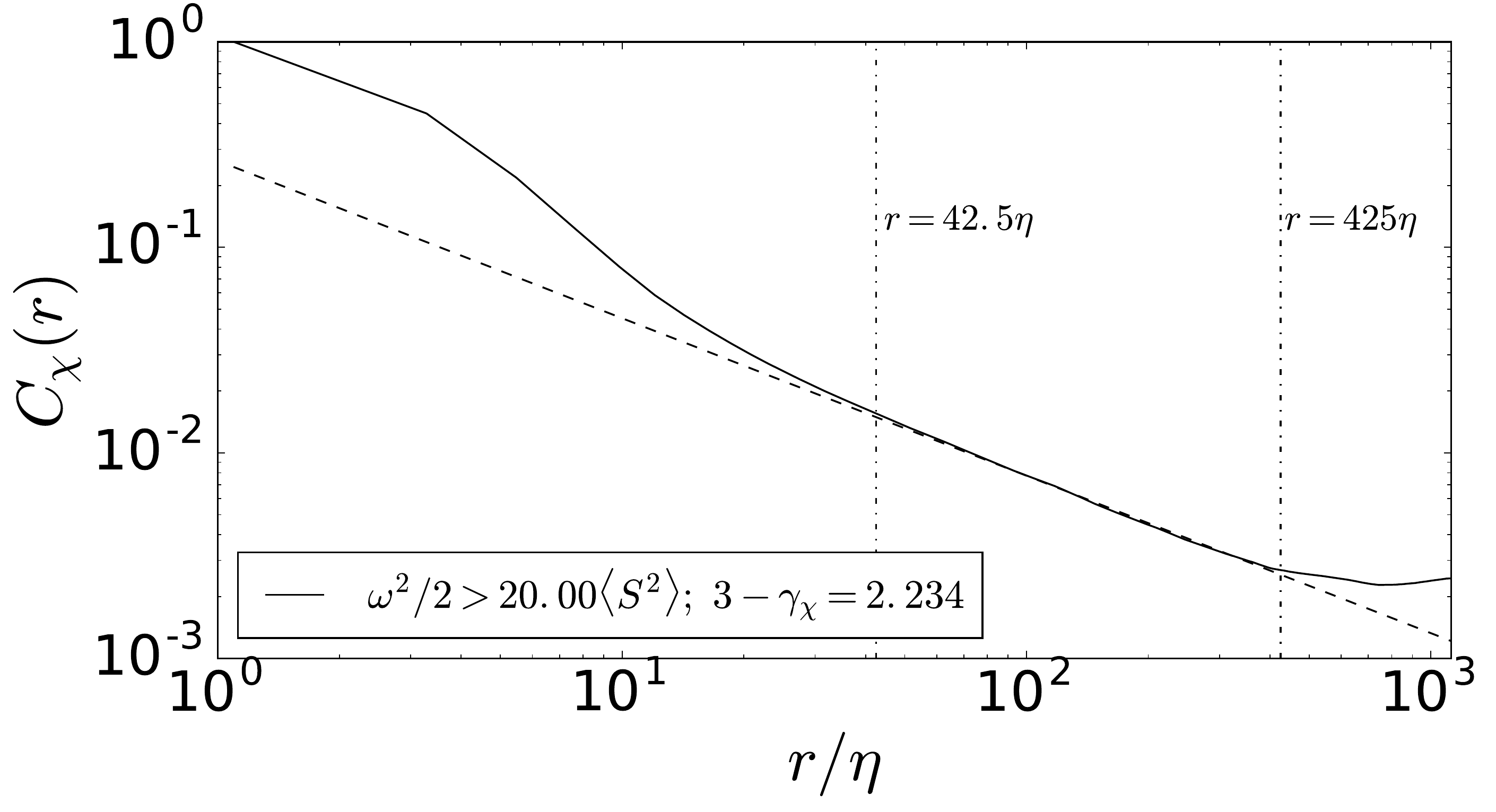}
            \put (-46,96){\makebox[0.05\linewidth][r]{(a)}}            
        \end{minipage}
        \begin{minipage}{0.49\linewidth}
            \includegraphics[width=\linewidth]{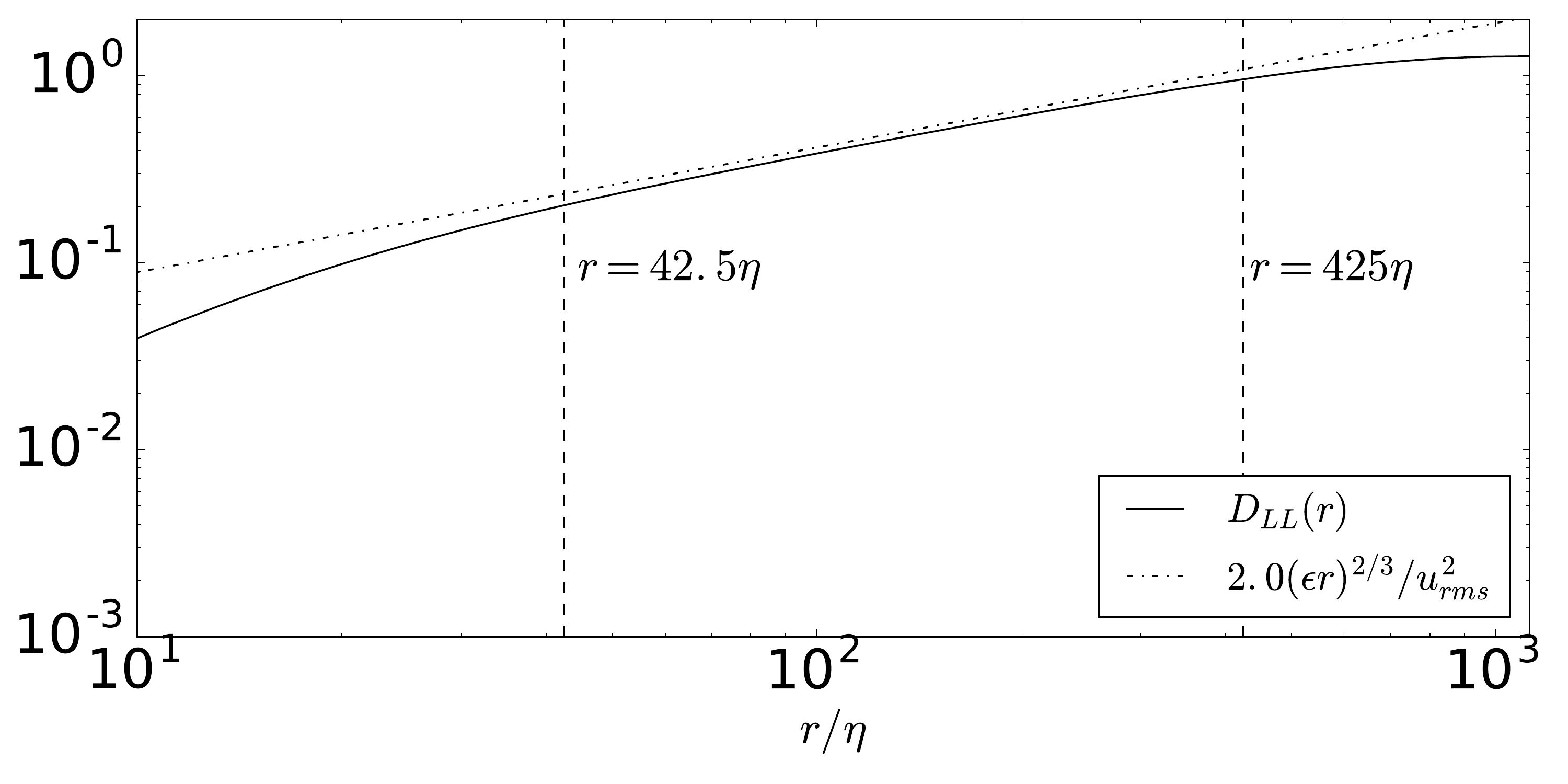}
            \put (-162,96){\makebox[0.05\linewidth][r]{(b)}}            
        \end{minipage}
        
        \caption{(a): Radial two-point correlation  function for the  $\omega^2/2 > 20\langle S^2\rangle$ excursion set. 
                 (b): Second-order longitudinal velocity structure function $D_{LL}$ as function of distance $r$. The dotted line is 
                 the classical Kolmogorov prediction for the longitudinal $2^{\rm nd}$-order structure function with $C_1=2.0$ (the small difference with the
                 leading coefficient that is usually closer to $C_1 \approx 2.1$ can be attributed to statistical convergence since we are only using a single snapshot).}
        \label{fig:excursion-20ab}
      \end{figure}

      The radial two-point correlation function of the enstrophy excursion set  corresponding to $\chi={20\langle S^2\rangle}$ is plotted in log-log axes in figure \ref{fig:excursion-20ab}(a). 
      A power-law tail is clearly visible over about a decade, between $42.5\ \eta < r < 425\ \eta$. This range  corresponds, roughly, to inertial range 
      of the studied dataset.  For direct comparison, we  computed the  second-order structure functions for the dataset, as shown in figure  \ref{fig:excursion-20ab}(b).
    The structure function has a scaling exponent of about $\xi_2 = 0.68$ which is the known value (slightly above the K41 value of 2/3 due to intermittency, \cite{1995Frisch}). 
    These plots allow us to compare the quality and range of the power-laws found in both observables. The most important fact to notice is that the 
    range where the two-point correlation function exhibits a near power-law behavior (the interval $42.5\ \eta < r < 425\ \eta$) is the same as in the structure function.    
    A similar behavior will be observed for all the excursion, iso-sets, and joint distribution sets studied in this work.
                
      The scaling exponent observed, in figure \ref{fig:excursion-20ab} for the enstrophy excursion set at the given threshold is about 
    $\gamma_\chi \approx 0.766$, implying a ``correlation dimension'' of about $D(\chi) \approx 2.234$. Thus, while the topology 
    of each individual vortex structure is visibly more one-dimensional, as a set its two-point structure is significantly more ``space filling'' 
    with a correlation structure that decays  more slowly on average than a collection of isolated vortices. 

          It is important to note that correlation functions were evaluated for the indicator function distribution directly (i.e. a field of ones and zeros), and not the ``fluctuation'' of 
          the indicator function away from its spatial mean which would include negative values by necessity.  
          We also tried to perform calculations on the subtracted version of the correlation function,  
          but the resulting correlation functions do not present as clear a power-law behavior in the inertial range as the one without subtracting the mean. 
          One plausibility argument for this observation is that the correlation function without subtracting the mean more readily corresponds to the definition of the mass dimension in which the mass in 
          spheres of radius $r$ is evaluated, and scaling with distance $r$ is used to define the mass dimension 
          \cite{mandelbrot1982fractal}
      
      \subsection{Box-counting based dimensions}
      
    As an independent measure of fractal dimension for the excursion sets we can also compute the box-counting dimension and the 
    box-counting based correlation dimension. The box-counting procedure for evaluating both of these dimensions
    is based on a set of cubes $B_{r, \bf k}$ of size $r$ and location identified by indices ${\bf k} = (k_1,k_2,k_3)$ so that a cube's corner is located at $[k_1 r, k_2 r, k_3 r]$ 
    with $k_1,k_2,k_3 \in \mathbb{Z}$, and $0 \leq k_i \leq \lceil 2 \pi/r\rceil$. We assign a measure to each cube, given by
      
      \begin{equation}    
       \mu_\chi(B_{r,\bf k}) = \frac{1}{V_{2\pi}} \int_{B_{r, \bf k}} \Theta_\chi({\bf x}) d^3{\bf x}, ~~ {\rm where} ~~ V_{2\pi} = \int_{[0,2\pi]^3} \Theta_\chi({\bf x}) d^3{\bf x}.
      \end{equation}
     
       The scaling of $N_r$, the number of boxes needed to cover the set, and of  $\sum_{{\bf k}} \mu_\chi(B_{r,{\bf k}})^2$ is 
     used to define the box-counting dimension and the box-counting based correlation dimension, respectively. 
     We thus compute
      
      \begin{equation}
        N_r =  \sum_{\bf k} [\mu_\chi(B_{r, {\bf k}})]^0, ~{\rm where} ~
                                      \ \ ~ \mu_\chi(B_{r, \bf k})^0 = \begin{cases}
                                                                    1, & \text{if } \int_{B_{r, {\bf k}}} \Theta_\chi({\bf x}) d^3{\bf x} > 0  \\
                                                                    0, & \text{otherwise}
                                                                   \end{cases}
      \end{equation}
      as well as 
      
      \begin{equation}
        {\cal M}_{2}(r)  =  \sum_{\bf k} [\mu_\chi(B_{r,{\bf k}})]^2.
      \end{equation}
     
       The behavior  $N_r \sim (r/\eta)^{-D_0}$ defines the box-counting dimension $D_0$, and
       ${\cal M}_{2}(r) \sim (r/\eta)^{D_2}$ defines the box-counting based correlation dimension, $D_2$ 
       \cite{Hentschel1983infinite,MeneveauSreenivasanJFM91}.
       
       The implementation of the box-counting dimension is done as follows: The positions of all points in the set are 
     histogrammed using the cubic box boundaries as the bins boundaries. For each bin with non-zero count, the bin was 
     normalized to $1$, and all other bins are left to $0$. The resulting histogram is summed, yielding the number of boxes
     that intersect the set of interest, $N_r$. The box size $r$ ranges between $4.25\ \eta$ and $850\ \eta$. 
     For the box-counting based correlation dimension, the computation is similar, but instead of normalizing the resulting count,
     we compute the sum of the bin values squared, which amounts to the $\mu_\chi(B_{r,k})^2$ calculation. 
       
\begin{figure}[H]
          \begin{minipage}{0.49\linewidth}
            \begin{center}
            \includegraphics[width=0.97\linewidth]{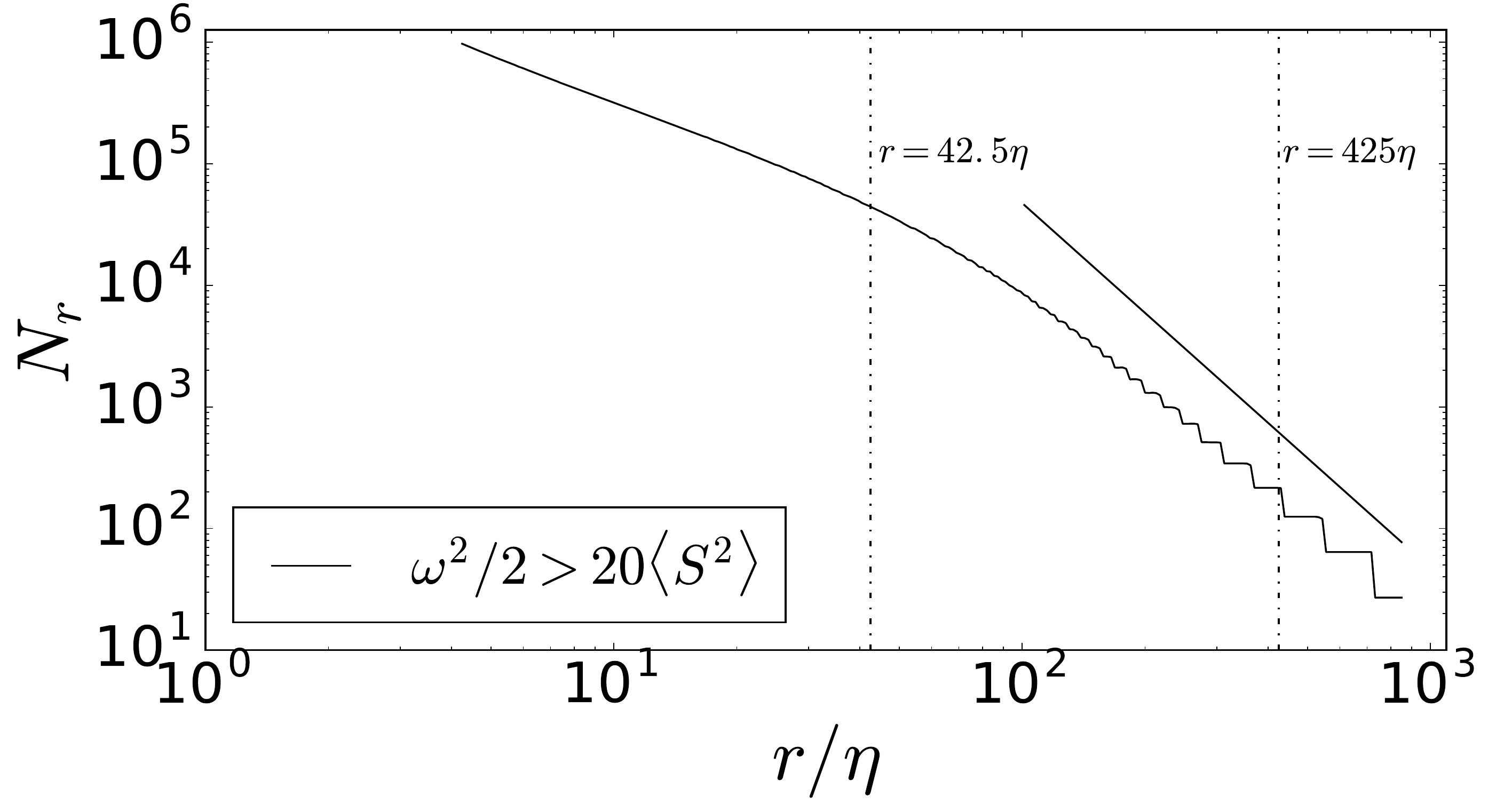}
            \put (-190,106){\makebox[0.05\linewidth][r]{(a)}}
            \end{center}

          \end{minipage}
          \begin{minipage}{0.49\linewidth}
            \begin{center}
            \includegraphics[width=\linewidth]{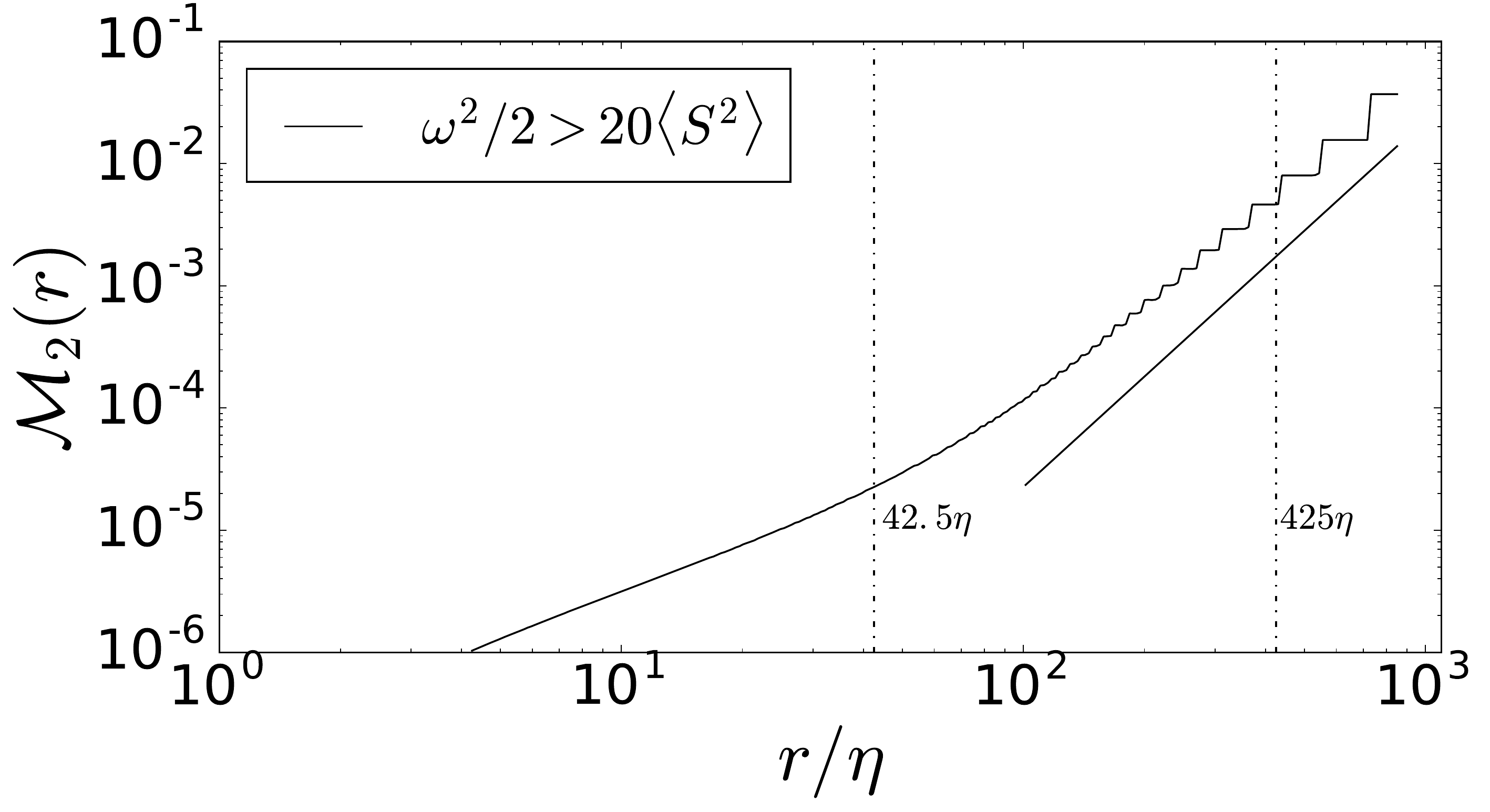}
            \put (-190,81){\makebox[0.05\linewidth][r]{(b)}}             
            \end{center}

          \end{minipage}
        
        \caption{(a): Box-counting dimension plot for the enstrophy excursion set with threshold
                 $\chi = 20\langle S^2\rangle$. (b): Box-counting  based correlation dimension plot for the 
                 same set. The black line represents the $D_0 = D_2 \sim 3$ scaling expected for the large scales of the box-counting 
                 calculation, since at large scales the clusters appear space-filling.}
        \label{fig:str-corr}
      \end{figure}

       The results for the box-counting dimension can be seen in Fig. \ref{fig:str-corr}(a)
     and the box-counting based correlation dimension plot is presented in Fig. \ref{fig:str-corr}(b). The most notable feature 
     of both plots is the lack of an inertial range scaling behavior, in contrast with the structure function and the correlation 
     dimension for the same set, Fig. \ref{fig:excursion-20ab}(a). One can discern a scaling at small scales $r < 20 \eta$ in 
     Fig.  \ref{fig:str-corr}(a) with a slope near -1 which implies $D_0 \sim 1$, i.e. 1-D objects, not unexpected for vortices 
     at the viscous scales. However, no scaling is observed in the inertial range with smooth curving towards a slope of 
     -3 (space-filling) at larger scales $r>200 \eta$ approaching the integral scale. Similar conclusions are reached from 
     the plots in Fig. \ref{fig:str-corr}(b). Hence, no inertial-range power-law scaling is found for the box-counting approach 
     applied to the excursion set of enstrophy. We have verified that the same is true for dissipation and all other variables 
     considered in this paper (not shown but some limited results will be shown later). Hence, we focus our further analysis on
     the correlation-function based analysis and scaling. We have tested that the correlation and box-counting algorithms yield
     correct results based on a 3D fractal set of known dimension (the Menger sponge), as summarized in Appendix B.  
       
    \subsection{Dependence on threshold, and dissipation-based excursion sets}
             
\begin{figure}[H]
        \begin{minipage}{0.34\linewidth}
          \includegraphics[width=\linewidth]{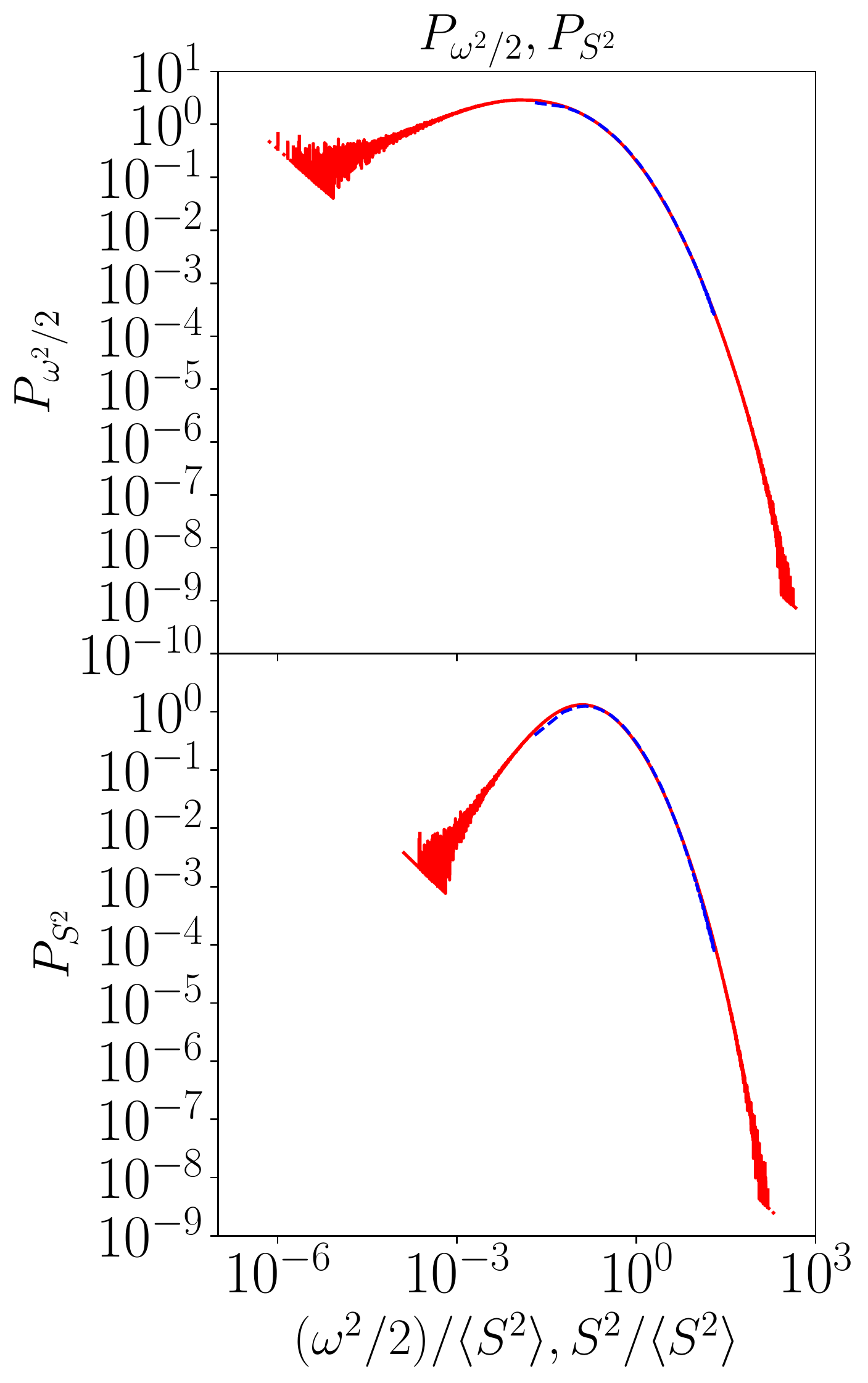}
          \put (-18,228){\makebox[0.05\linewidth][r]{\Large (a)}}
          \put (-18,121){\makebox[0.05\linewidth][r]{\Large (b)}}
        \end{minipage}
        \begin{minipage}{0.65\linewidth}
          \includegraphics[width=\linewidth]{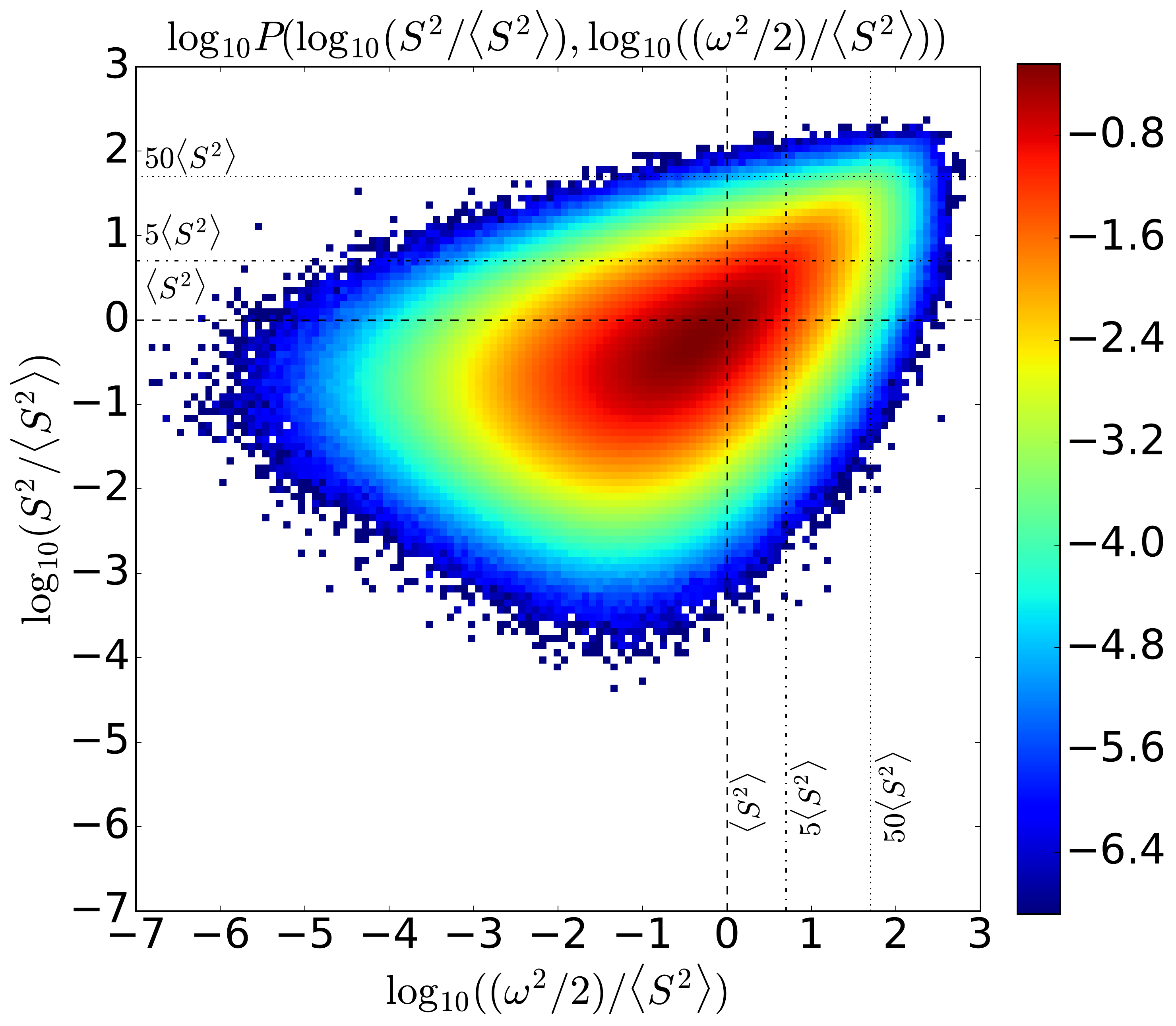}
          \put ( -66,238){\makebox[0.05\linewidth][r]{\Large (c)}}
        \end{minipage}
          
        \caption{(a) Enstrophy and (b) Dissipation PDFs, computed from DNS of forced isotropic turbulence at $Re_\lambda \sim 430$. 
                 The dashed blue line is the same PDF computed using a different analysis program for this same dataset as 
                 reported in Ref. \cite{johnson2015large}. 
                 (c) Enstrophy and Dissipation Joint-PDF. A
                 similar joint PDF can be found in reference \cite{borue_orszag_1998}. 
                 }
        \label{fig:enst-strain-pdf}
      \end{figure}

\begin{figure}[H]        
        \begin{center}
        \begin{minipage}{0.49\linewidth}
          \includegraphics[width=\linewidth]{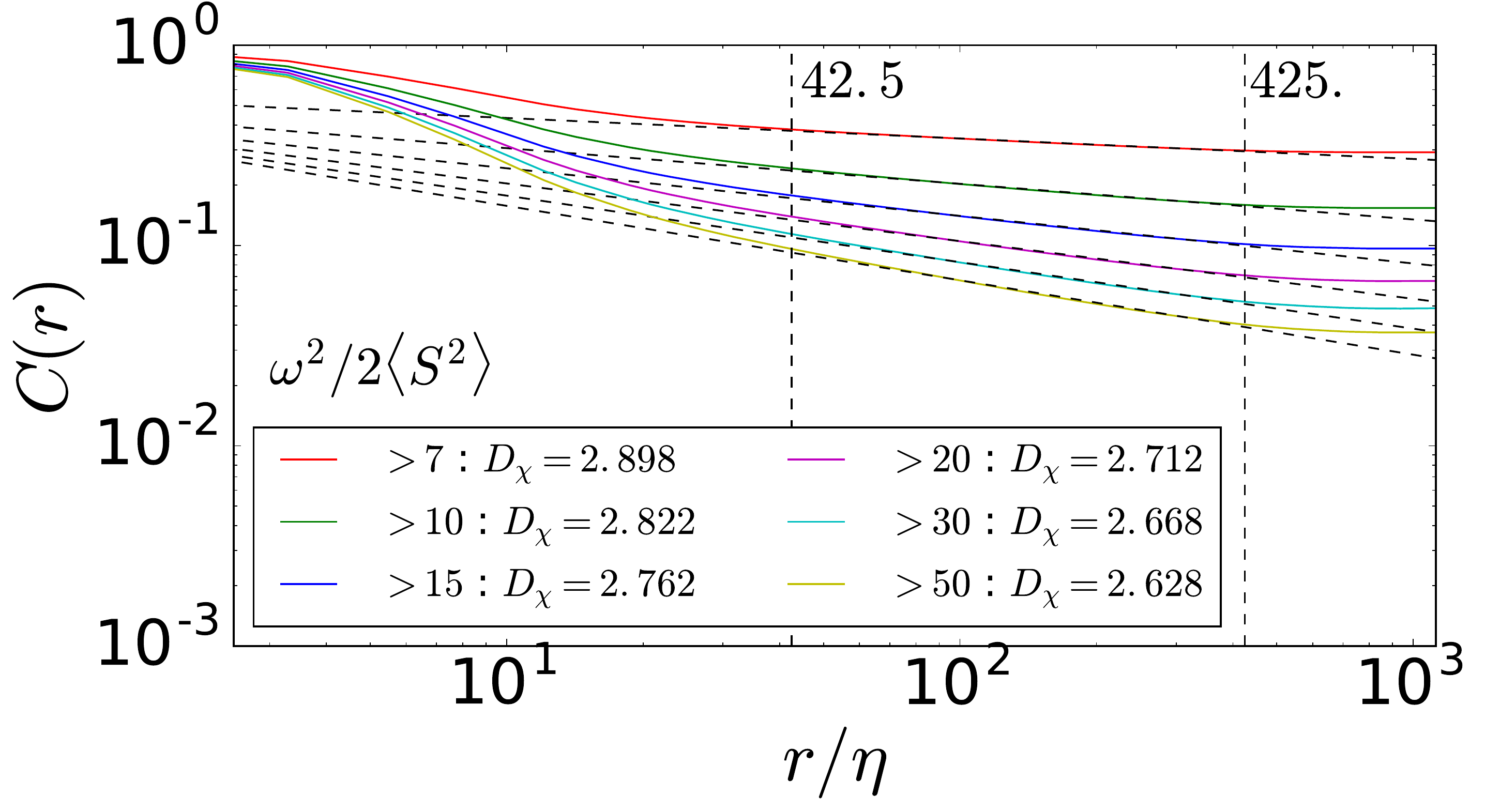}
          \put (-185,85){\makebox[0.05\linewidth][r]{\Large (a)}}
        \end{minipage}
        \begin{minipage}{0.49\linewidth}
          \includegraphics[width=\linewidth]{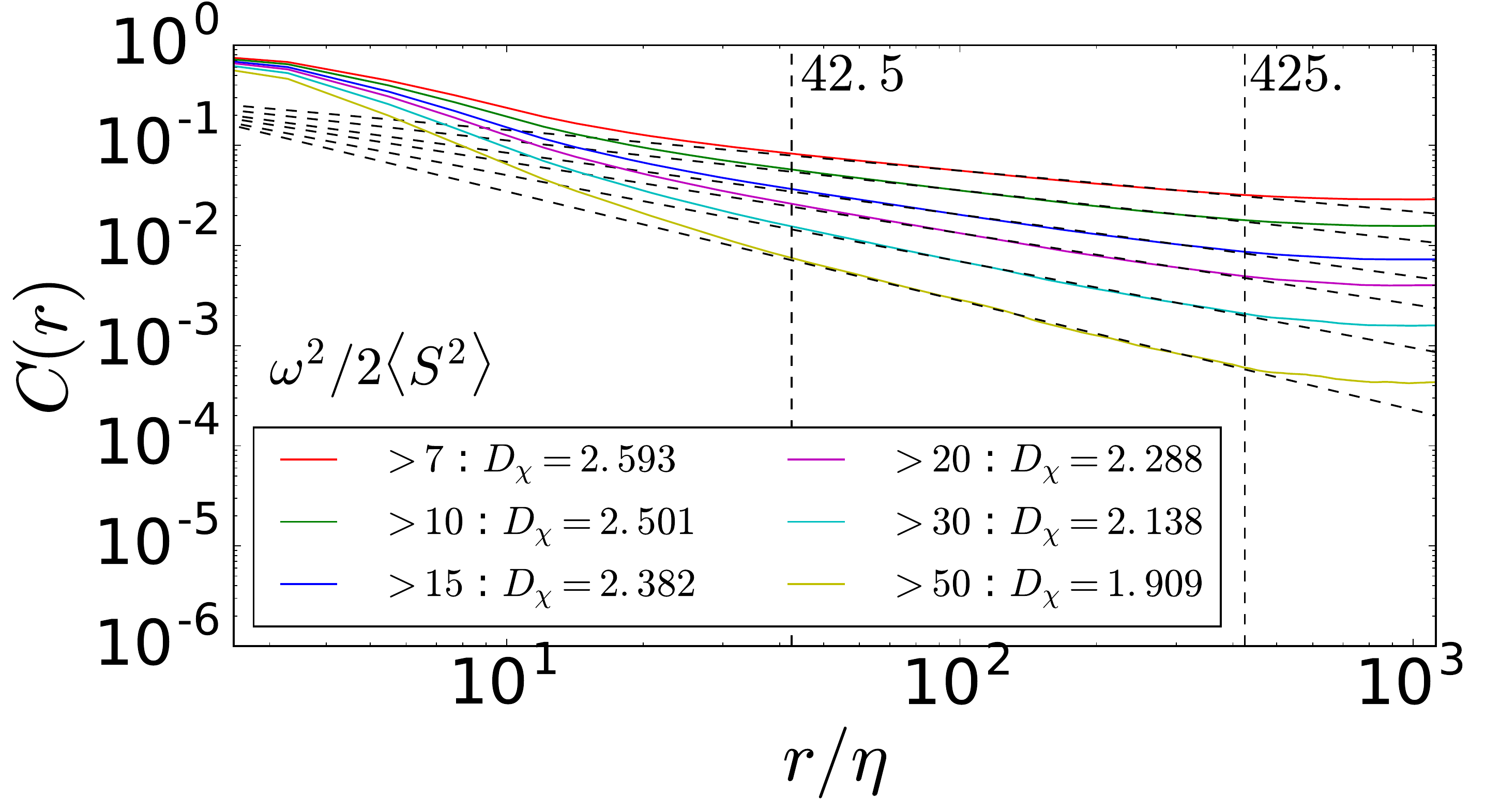}
          \put (-185,90){\makebox[0.05\linewidth][r]{\Large (b)}}
        \end{minipage}
      
        \begin{minipage}{0.49\linewidth}
          \includegraphics[width=\linewidth]{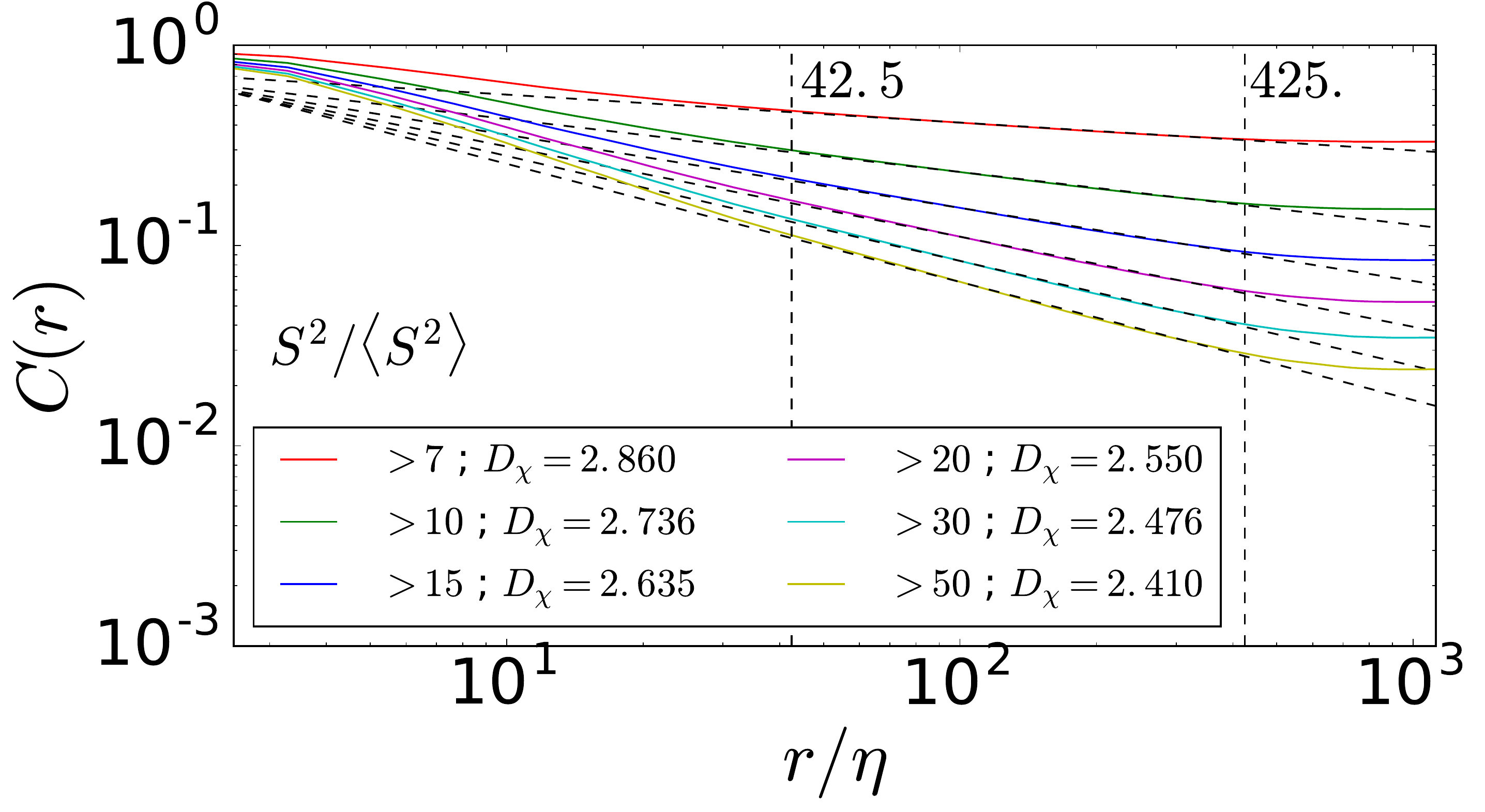}
          \put (-185,90){\makebox[0.05\linewidth][r]{\Large (c)}}
        \end{minipage}
        \begin{minipage}{0.49\linewidth}
          \includegraphics[width=\linewidth]{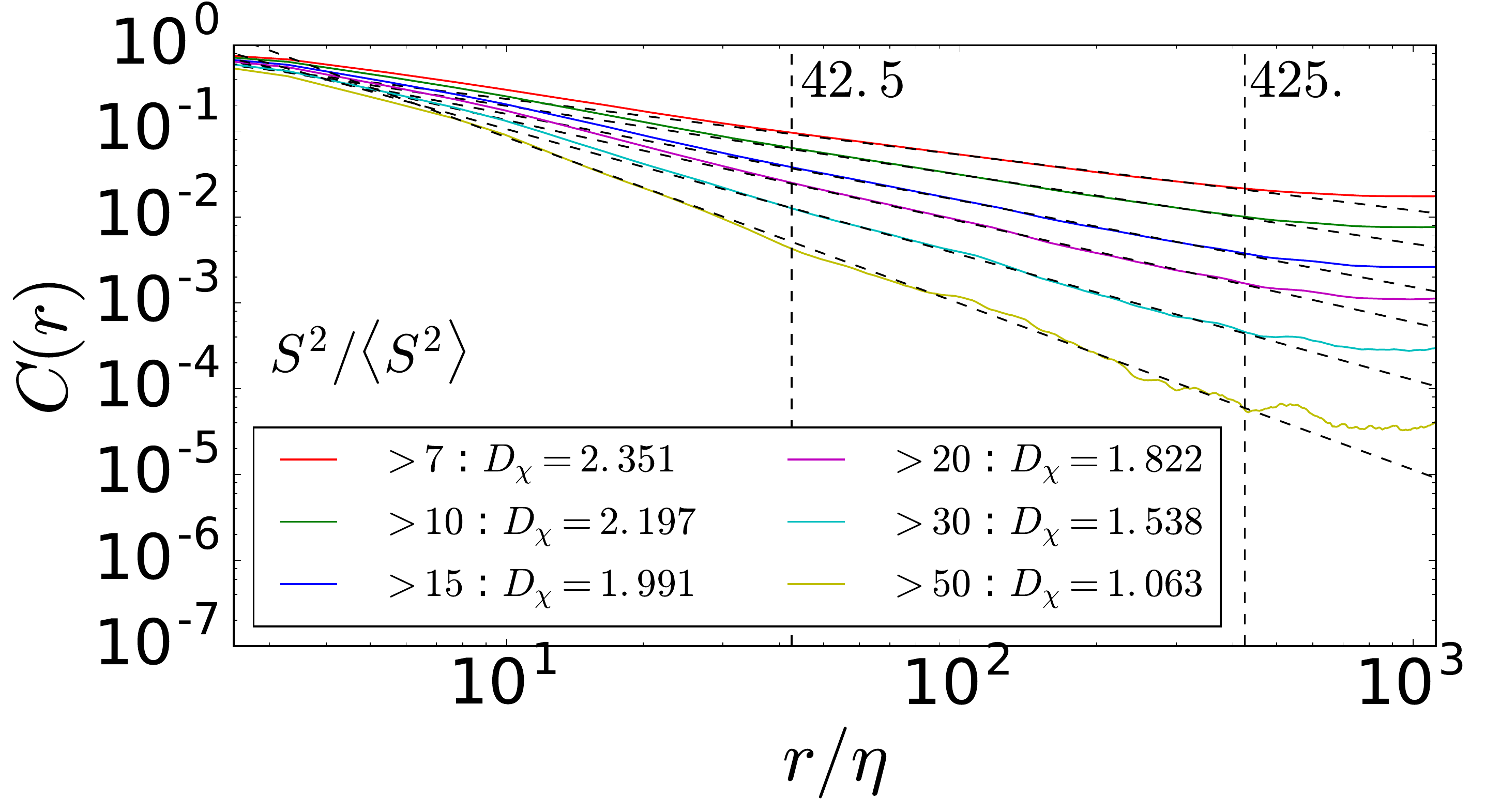}
          \put (-185,90){\makebox[0.05\linewidth][r]{\Large (d)}}
        \end{minipage}
        \end{center}
        
        \caption{Radial correlation functions for enstrophy $\omega^2/2$ (a,b) and dissipation $S^2$ (c,d) computed from DNS 
                 of forced isotropic turbulence at $Re_\lambda \sim 430$. The thresholds $\chi$ range from 
                 $1\langle S^2\rangle$ to $50\langle S^2\rangle$. The legend denotes $> \chi/\langle S^2\rangle$ threshold 
                 for each line.}  
        \label{fig:corr-enstr-strain-fam}    
      \end{figure}

      In this section we examine the correlation function scaling as a function of  threshold and also present a similar analysis for the 
    dissipation $S^2$. To place the thresholds in proper context,  in figure \eqref{fig:enst-strain-pdf} we present the 
    probability density function (PDF) for both enstrophy and dissipation, together with their joint  PDF, for the dataset we 
    used.  Varying the threshold $\chi$ we can probe different intensities of events, therefore different sectors of the PDFs. We 
    present, in figure (\ref{fig:corr-enstr-strain-fam}), the results for the correlation functions for several
    thresholds, ranging from $\chi = 1\langle S^2\rangle$ to $\chi = 50\langle S^2\rangle$, for both enstrophy (see Figs.
    \ref{fig:corr-enstr-strain-fam}(a,b)) and for dissipation (Fig. \ref{fig:corr-enstr-strain-fam}(c,d)).
    
     Clearly  all correlation functions shown in Figure \ref{fig:corr-enstr-strain-fam} present  power-law behavior within 
    the inertial range regardless of the observable probed and the value of the threshold. 
    As expected, the correlation dimension has a monotonically decreasing behavior 
    as a function of the threshold, indicating that high intensity sets become less and less space filling. 
    
     Comparing the values presented in Figures \ref{fig:corr-enstr-strain-fam}(a,c) with \ref{fig:corr-enstr-strain-fam}(b,d)
    confirms the initial observations made about figure \ref{fig:excursion-20-vol}, in which is clear that the fractal 
    dimensions associated with dissipation sets are systematically lower than the ones of the enstrophy sets, for the same 
    thresholding value. 
    
      This behavior is, initially, counterintuitive due to the expectation that enstrophy should be distributed along tubes, 
    i.e. elongated one-dimensional sets, while dissipation should be distributed along sheets. This behavior  is expected
    on the smallest, viscous, scales but does not seem  not to be reflected  in the inertial range behaviour, at least not 
    in the correlation function scaling. 
         
      Another feature visible in by Fig. \ref{fig:excursion-20-vol} is that enstrophy and strainrate are quite highly correlated,
    which can also be infered from the overall shape of the joint PDF shown in Fig. \eqref{fig:enst-strain-pdf}(c). 
    Quantitatively, we confirmed this by computing the correlation coefficients as follows:
    
      \begin{eqnarray}
         \rho(S^2,\omega^2/2) &=& \frac{\left\langle (S^2 - \langle S^2\rangle)(\omega^2/2 - \langle S^2\rangle)\right\rangle}{ {\langle (S^2 - \langle S^2 \rangle)^2 \rangle^{1/2} \langle (\omega^2/2 - \langle S^2 \rangle)^2 \rangle^{1/2}} } = 0.615 .\\
         \rho(\log S^2,\log \omega^2/2) &=& \frac{\left\langle (\log S^2 /\langle S^2\rangle)(\log \omega^2/2\langle S^2\rangle)\right\rangle}{ {\langle (\log S^2/\langle S^2 \rangle)^2 \rangle^{1/2} \langle (\log \omega^2/2\langle S^2 \rangle)^2 \rangle}^{1/2} } = 0.675 .
      \end{eqnarray}
      
      The non-negligible correlation between enstrophy and dissipation has been noted before, see, e.g. 
    \cite{Zhu1996,donzisYeungSree2008,gualaLiberzonTsinoberKinzelbach2007}.
      
\begin{figure}[H]
        \begin{minipage}{0.49\linewidth}
        \includegraphics[width=\linewidth]{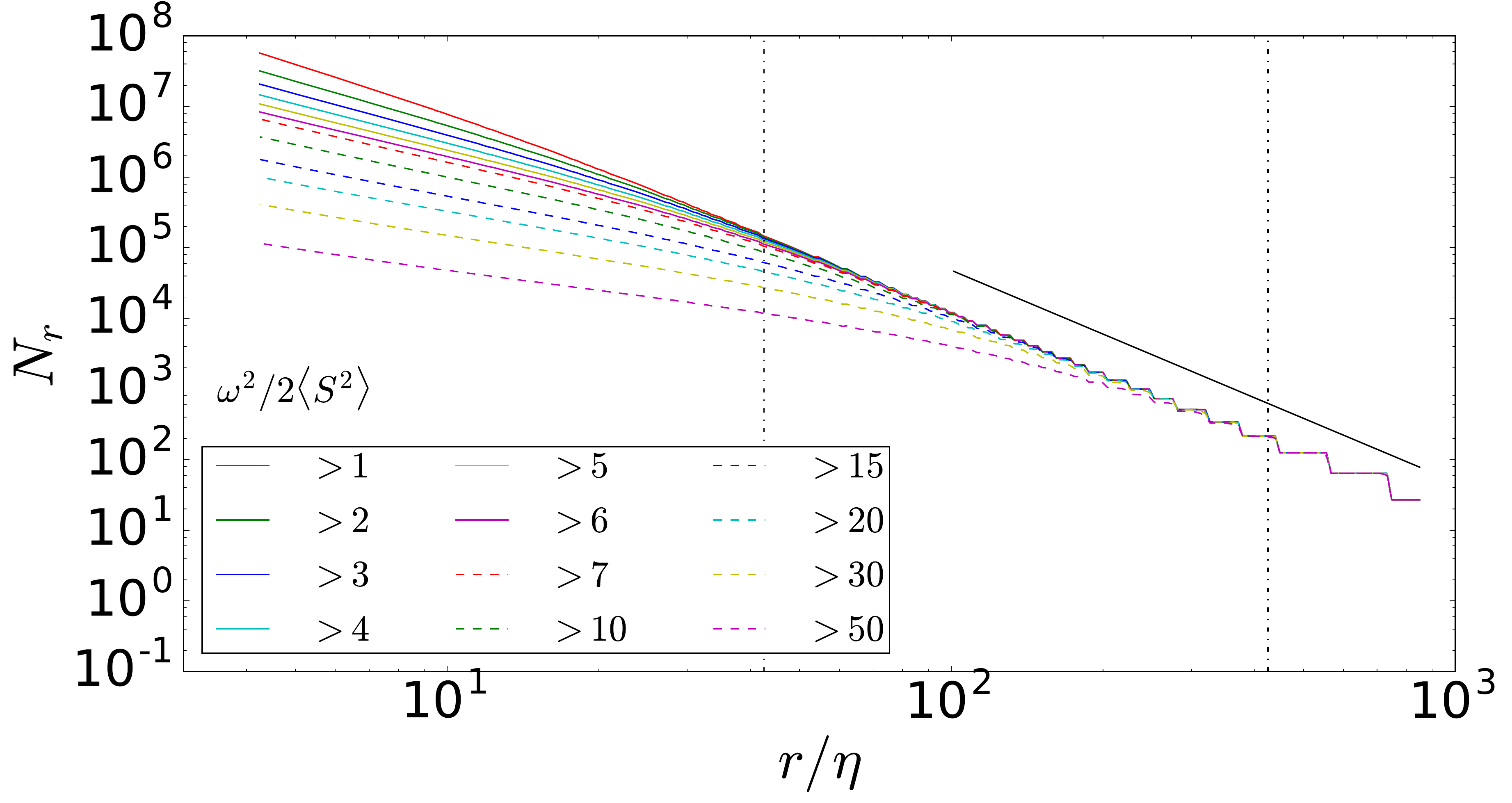}
          \put (-23,99){\makebox[0.05\linewidth][r]{\Large (a)}}
        \end{minipage}
        \begin{minipage}{0.49\linewidth}
        \includegraphics[width=\linewidth]{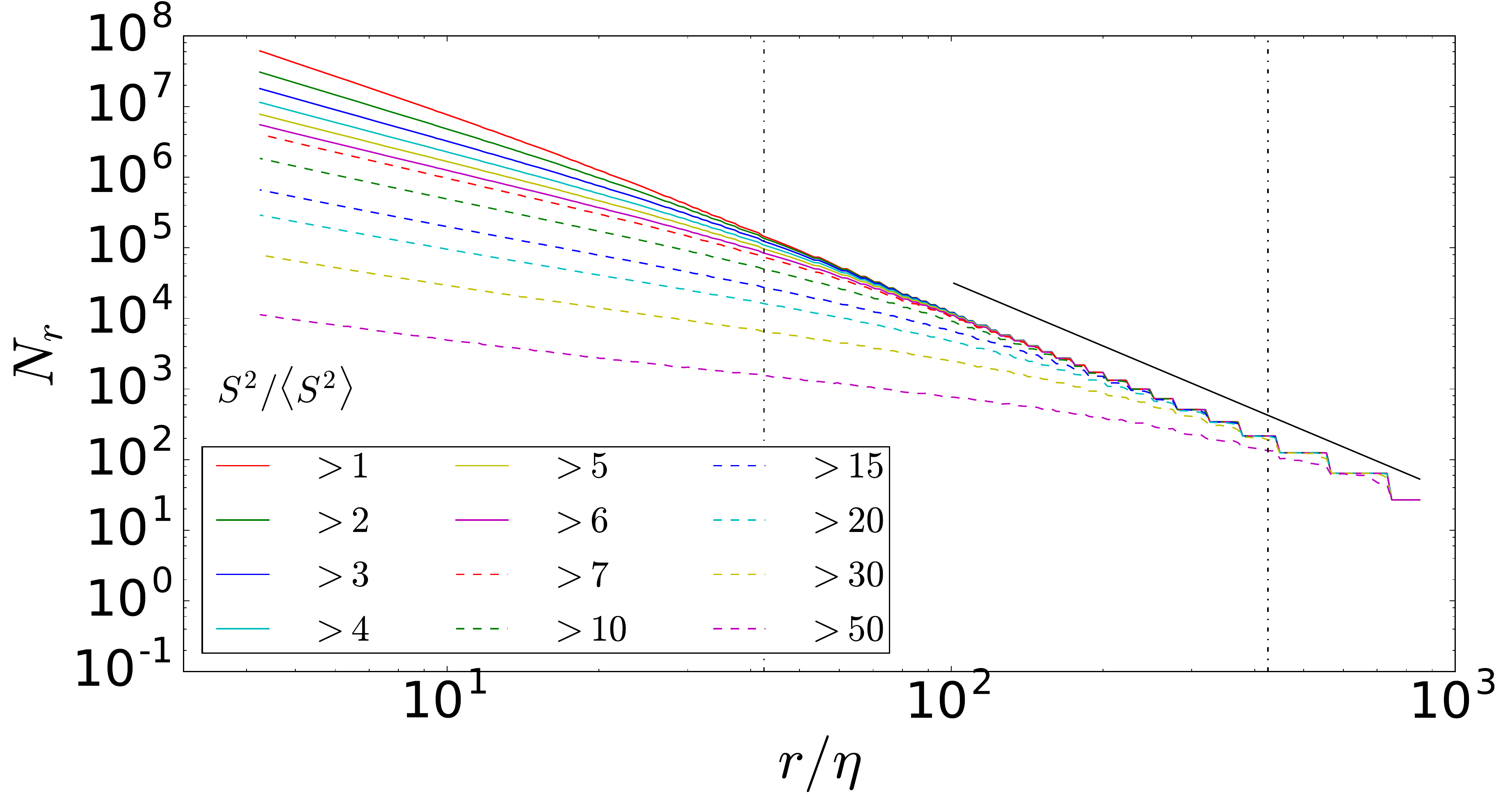}
          \put (-23,99){\makebox[0.05\linewidth][r]{\Large (b)}}
        \end{minipage}
         \caption{Box-counting dimension calculation for enstrophy (a) and dissipation (b) excursion sets showing 
                  lack of power-law scaling in the inertial range. The black line represents the $D_0 = 3$ scaling
                  expected for the large scales. The legend denotes $> \chi/\langle S^2\rangle$ threshold for each line.}
        \label{fig:enstr-boxcount}
      \end{figure}

      For comparison we present, in Figure \ref{fig:enstr-boxcount}, the box-counting dimension  for these same sets, based on 
    thresholding enstrophy and dissipation. They present the  same features as in figure \ref{fig:str-corr}(a,b). At 
    the smallest scales we can see that most high-enstrophy sets approach  a slope of -1 but very high thresholds lead to even 
    shallower (smaller-in-magnitude) slopes, consistent with broken up, less coherent vortex events. For the dissipation 
    structures, at small scales we see a range of slopes slopes between -2 for intermediate thresholds (consistent with sheets)
    but also decreasing continuously towards -1 and lower for higher thresholds. For the larger scales we notice again that 
    all excursion sets saturate the box-counting to a slope of $\sim -3$ due to the homogeneity (space-fillingness) of the 
    turbulence structures at the largest scales.
    
      The lack of scaling of the box-counting results in the inertial range makes connection of the correlation function based
    exponent $D(\chi)=3-\gamma_\chi$ with a fractal dimension not as clear as one would hope. Therefore, while from here on
    we will refer to $D(\chi) = 3-\gamma_\chi$ as the correlation dimension, we must keep  these limitations in mind.
    
  \section{Interval-based (iso) sets } \label{sec:interval}
  
      An interesting alternative to considering excursion sets is to compute interval-based sets, which corresponds to the sets
    in which the observable is between  $\chi_-$ and $\chi_+$, a lower and a upper threshold respectively. 
    These sets are  defined according to 
      \begin{equation}
        \Theta^A_{\chi_+,\chi_-}({\bf x}) = H(A({\bf x})-\chi_-)H(\chi_+-A({\bf x}))  = 
        					\begin{cases}
                                             1, & \text{if } \chi_- \leq A({\bf x}) \leq \chi_+  \\
                                             0, & \text{otherwise}
                                          \end{cases}
      \end{equation}
    They  correspond  to subsets of the excursion set near the lowest threshold, i.e. near the iso-threshold set bounding the 
    excursion set. In fact this procedure yields a good approximation for an ``iso-set' when $(\chi_+ - \chi_-)/\chi_- \ll 1$.  
          
     We compute  the two-point correlation function using $\chi_-$ varying in the same set as the 
    thresholds of the previous section, and $\chi_+ = 1.05 \chi_- = (1+\Delta)\chi_- $, which roughly corresponds to a thin 
    shell of the inner boundary of the excursion set of threshold $\chi_-$. We prefer a multiplicative, rather than 
    additive, relationship between $\chi_+$ and $\chi_-$ because it amounts to equally sized logarithmic bins. We also
    refer to these ``interval-based" sets as ``shells" sets.
    
\begin{figure}[H]
        \begin{minipage}{0.32\linewidth}
          \includegraphics[width=\linewidth]{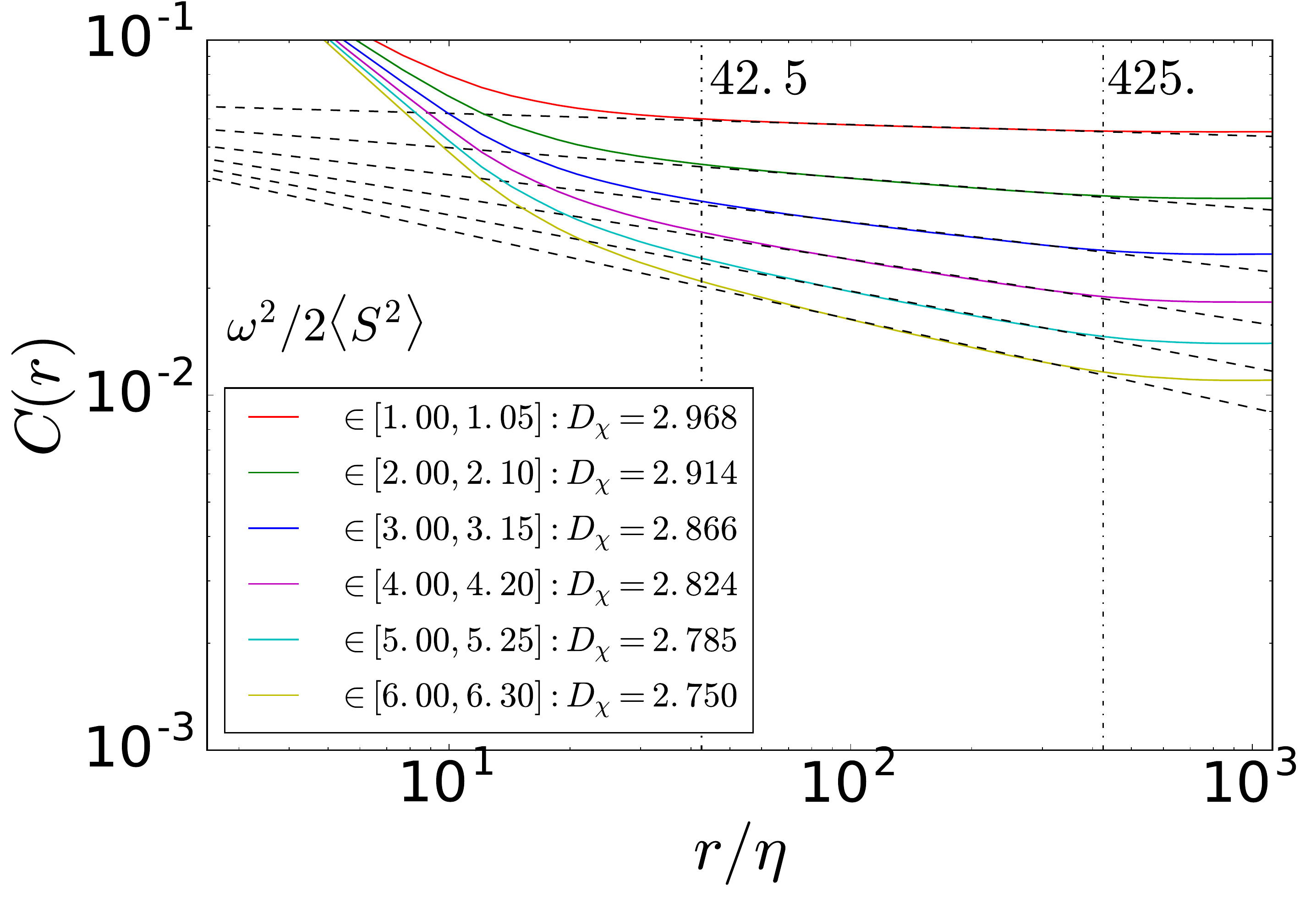}
          \put (-121,73){\makebox[0.05\linewidth][r]{(a)}}
        \end{minipage}
        \begin{minipage}{0.32\linewidth}
          \includegraphics[width=\linewidth]{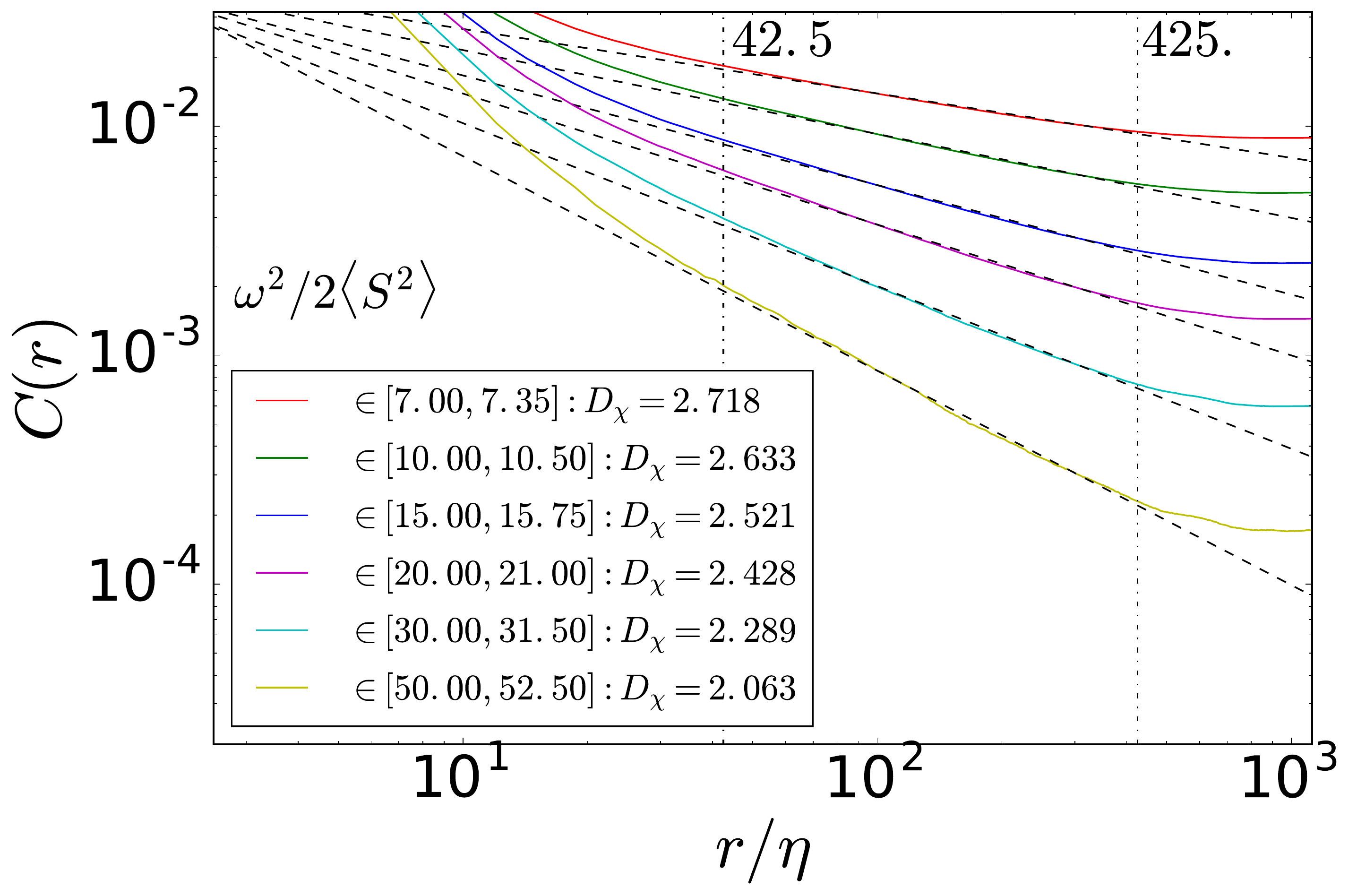}
          \put (-121,73){\makebox[0.05\linewidth][r]{(b)}}
        \end{minipage}
        \begin{minipage}{0.32\linewidth}
          \includegraphics[width=\linewidth]{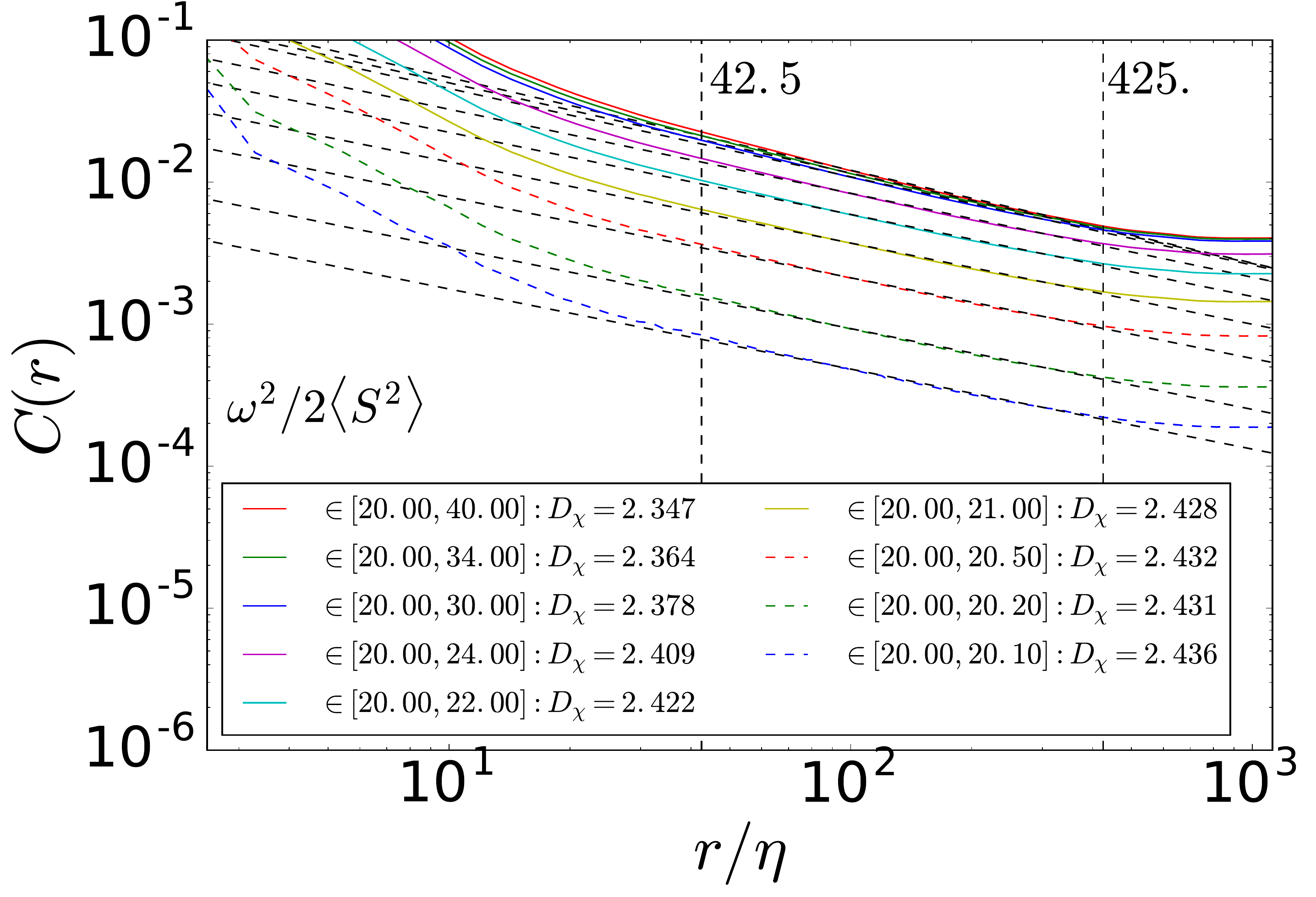}
          \put (-123,67){\makebox[0.05\linewidth][r]{(c)}}
        \end{minipage}   
        
        \begin{minipage}{0.32\linewidth}
          \includegraphics[width=\linewidth]{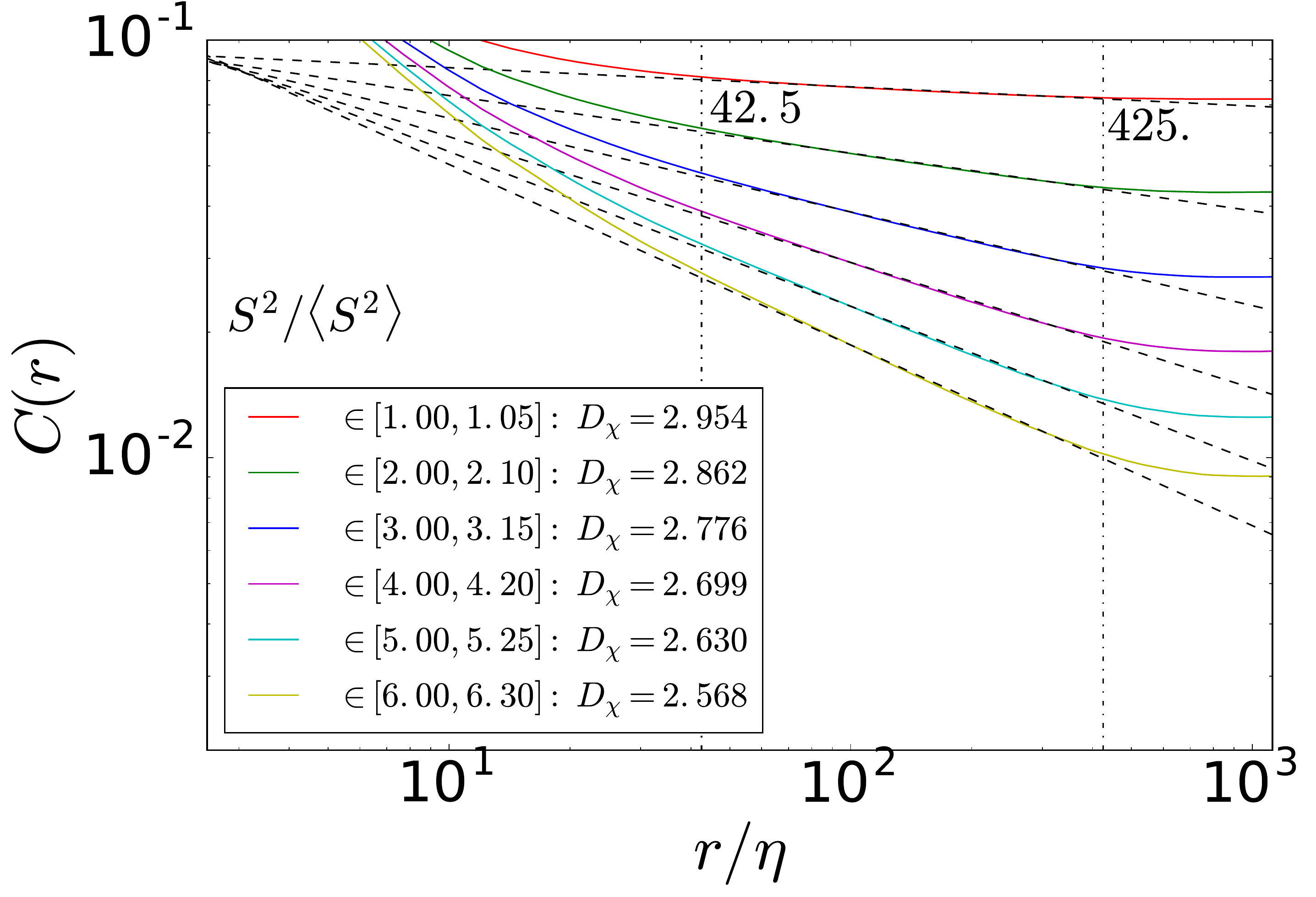}
          \put (-121,73){\makebox[0.05\linewidth][r]{(d)}}
        \end{minipage}
        \begin{minipage}{0.32\linewidth}
          \includegraphics[width=\linewidth]{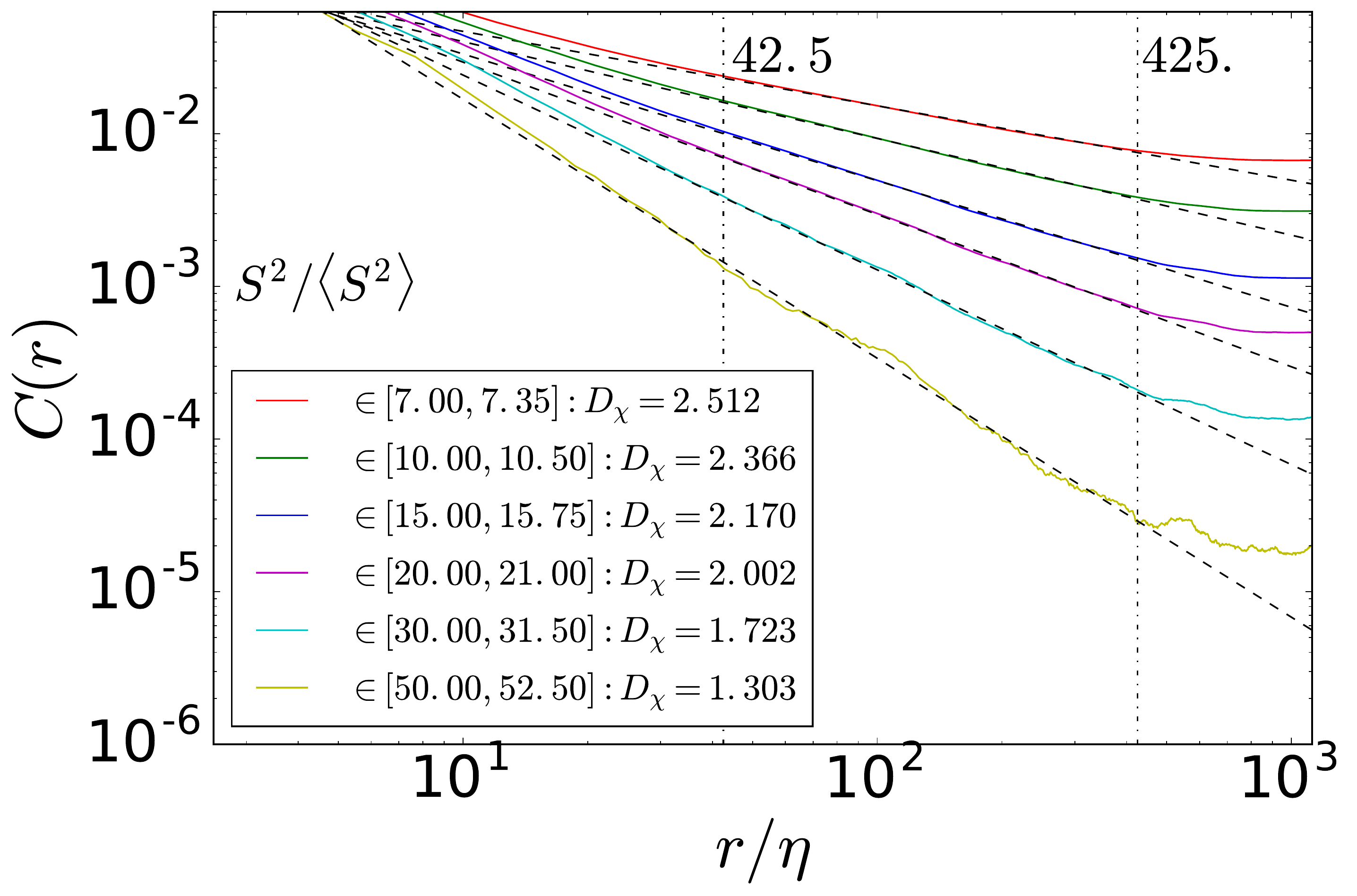}
          \put (-121,75){\makebox[0.05\linewidth][r]{(e)}}
        \end{minipage}
        \begin{minipage}{0.32\linewidth}
          \includegraphics[width=\linewidth]{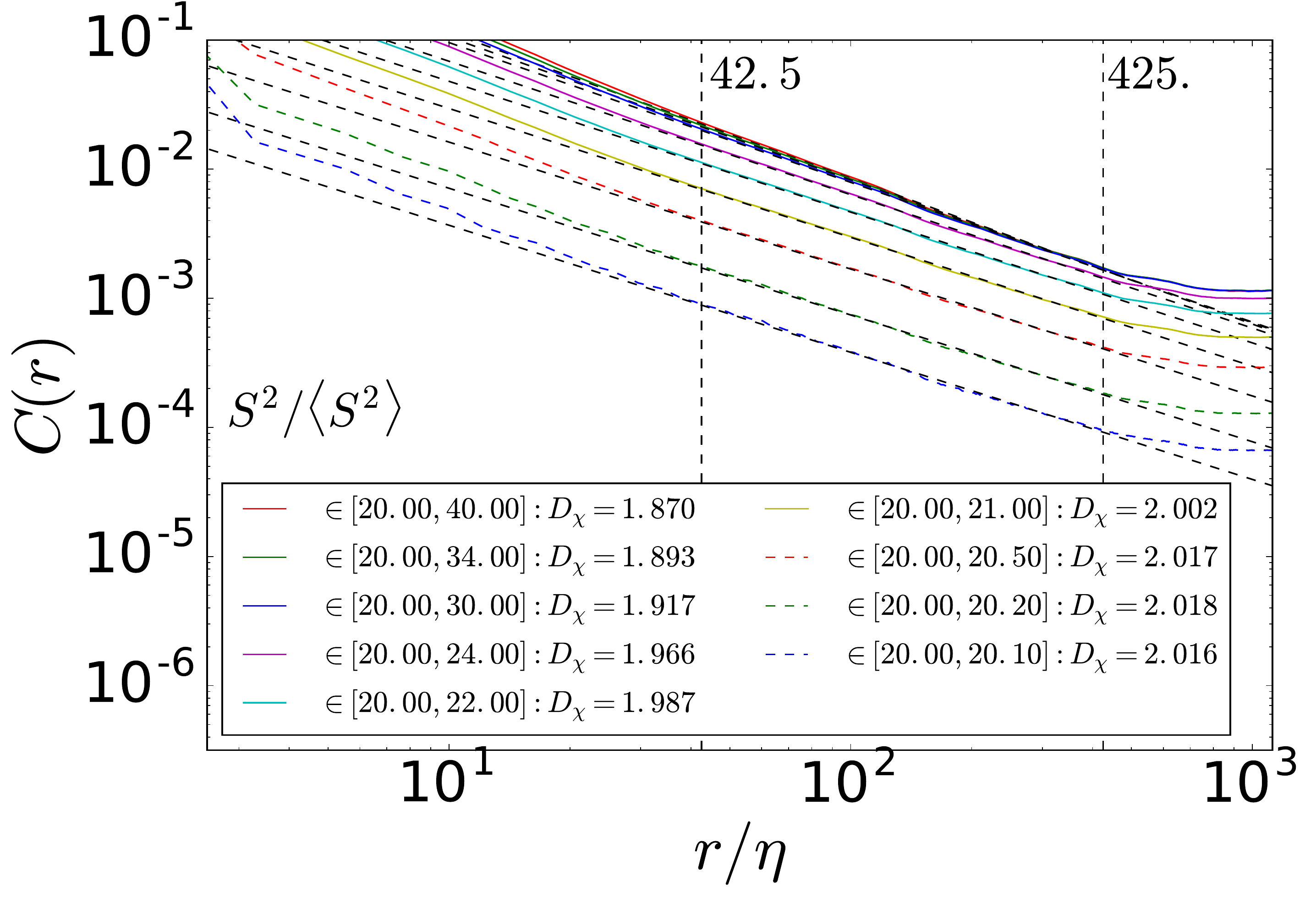}
          \put (-123,67){\makebox[0.05\linewidth][r]{(f)}}
        \end{minipage}
        \caption{Radial correlation functions for enstrophy $\omega^2/2$ and dissipation $S^2$ computed from DNS of forced isotropic 
                 turbulence at $Re_\lambda \sim 430$. The thresholds $\chi$ range from $1\langle S^2\rangle$ to 
                 $50\langle S^2\rangle$. The results are for shells of thickness $\chi_+-\chi_- = 0.05 \chi_-$, for 
                 both enstrophy, (a) and (b), and for dissipation, (d) and (e). As a band-width sensitivity test, we computed correlation 
                 functions for shells of varying thickness with base threshold $\chi_- = 20\langle S^2\rangle$, for both
                 enstrophy (c) and dissipation (f). Legend denotes $\in [\chi_-/\langle S^2\rangle,\chi_+/\langle S^2\rangle]$ 
                 interval for each observable.}
        \label{fig:corr-enstr-strain-diff}    
      \end{figure}

      The results of this analysis are presented in figure \ref{fig:corr-enstr-strain-diff}(a,b) for  enstrophy and 
    in figures  \ref{fig:corr-enstr-strain-diff}(c,d) for dissipation. As is visible, the resulting two-point correlations 
    also  present robust power-law scaling in the same inertial range as the previous excursion set correlation functions.
        
     In order to establish the robustness of results with regards to the ``thickness'' of the band of thresholds defining the 
   bin for the iso-set, we computed correlation functions for shells of varying thickness with base threshold 
   $\chi_- = 20\langle S^2\rangle$, for both enstrophy  and dissipation. As can be seen in Fig.  
   \ref{fig:corr-enstr-strain-diff}(c) for enstrophy, the resulting power-law can be observed to be robust regardless of the 
   tested thickness. Also,  the resulting exponent is insensitive to the thickness, unless very thick shells are used, in which 
   case we are actually closer to an excursion set than to a proper interval-based (iso) set. 
   A similar result can be obtained for the dissipation, as seen in Fig. \ref{fig:corr-enstr-strain-diff}(f).  
   Again, we repeated the box-counting dimensions computation for reference (not shown), and observed that  there is no  power-law in the 
   inertial range.     
   
      \begin{figure}[H]
        \begin{minipage}{0.46\linewidth}
          \includegraphics[width=\linewidth]{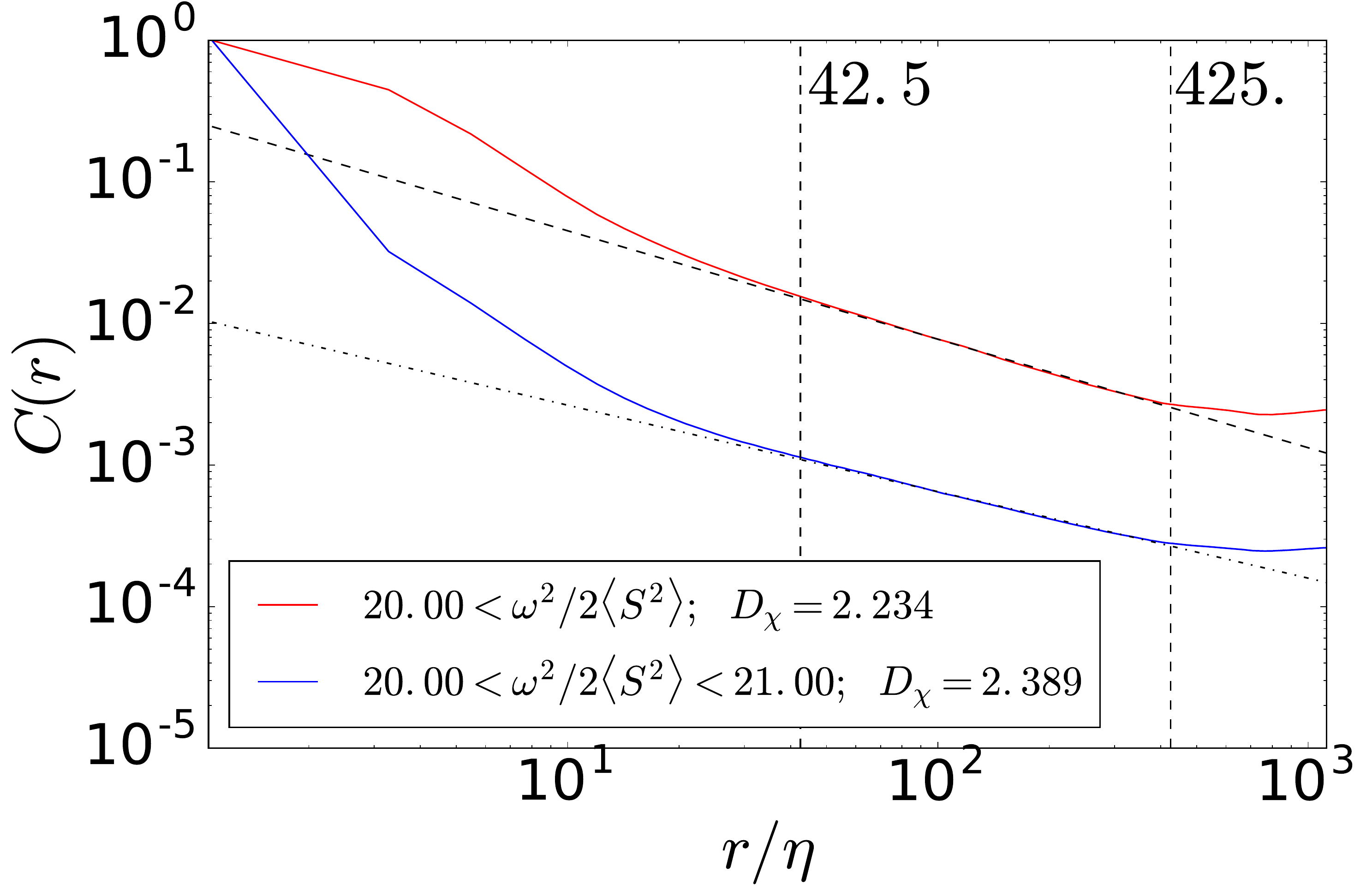}
          \put (-22,110){\makebox[0.05\linewidth][r]{\Large (a)}}
        \end{minipage}
        \begin{minipage}{0.53\linewidth}
          \includegraphics[width=\linewidth]{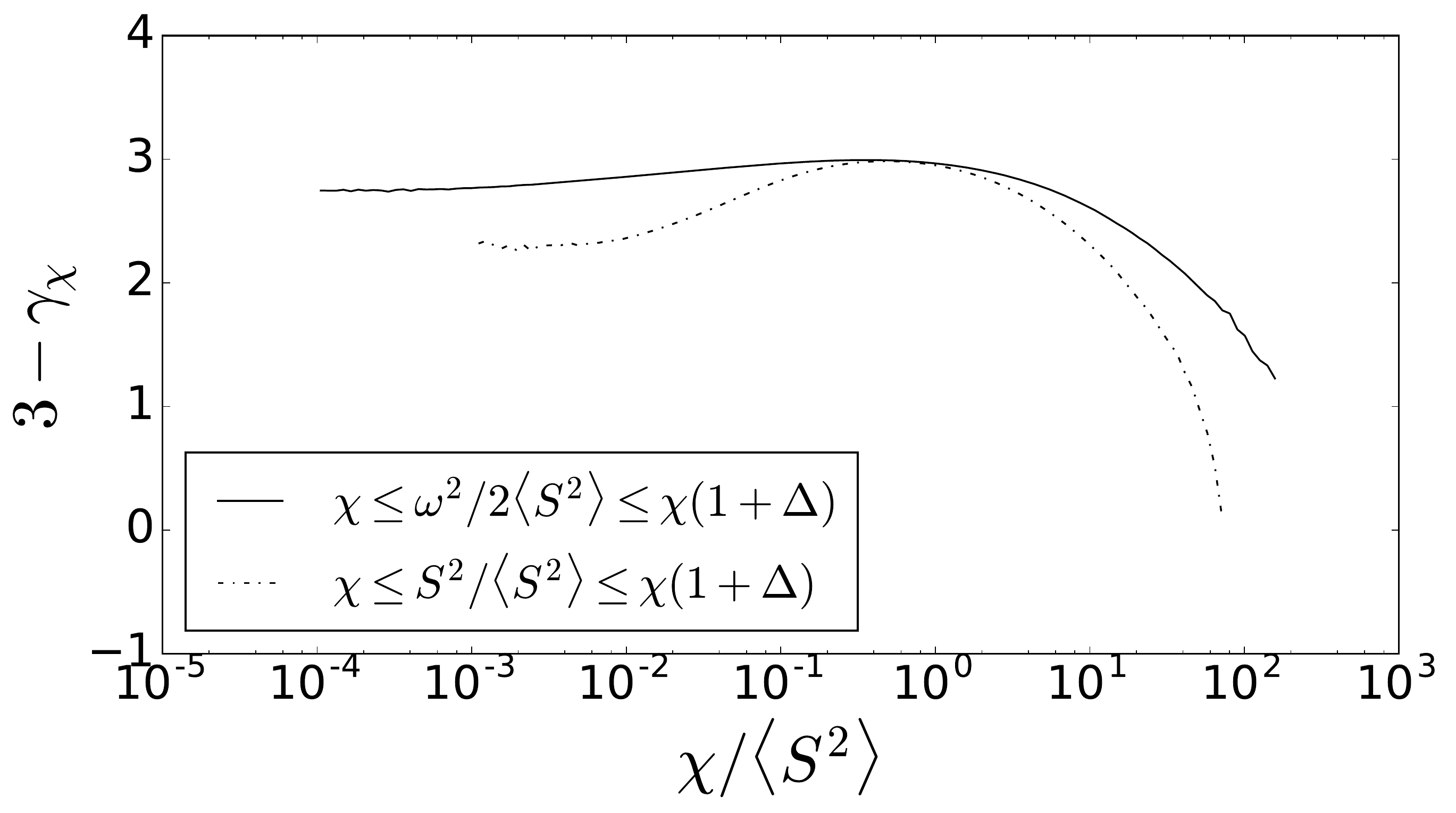}
          \put (-2,118){\makebox[0.05\linewidth][r]{\Large (b)}}
        \end{minipage}        
        \caption{(a) Comparison of excursion sets with thin shell sets starting at similar threshold values; (b) Interval-based sets dimensions, as a 
                 function of the probed intensity $\chi/\langle S^2\rangle$.}
        \label{fig:interval-dims}
      \end{figure} 
      
      To exemplify the difference in behavior between excursion sets and interval-based sets correlation functions, we
    computed both types of sets for the lower threshold  at $20\langle S^2\rangle$ in both cases.  One 
    observes  that they differ, mostly, on the small $r$ region, which is an imprint of the fact that shell-based sets have a 
     ``hollow" shape compared to the excursion sets. The absolute value of the normalized correlation function is lower for
    shells than for the respective excursion sets,  due to lower volume fraction, as expected. The 
    correlation slope is flatter, having a higher compensated exponent (i.e. more ``space-filling'').
    
    
      In order to present the complete information of scaling exponents as function of threshold  we present 
    $D(\chi) = 3-\gamma_{\chi}$ as function of threshold corresponding to logarithmically spaced bins.  The 
    computation is done for both enstrophy and dissipation shell sets. Overall both observables present the same qualitative 
    behavior, but enstrophy consistently shows a  higher correlation-based dimension. Prior results \cite{Meneveauetal90} have shown that enstrophy 
    is ``more intermittent'' than dissipation and thus the present results may appear to be counterintuitive,  as mentioned
    before in section \ref{sec:excursion}.  Present results show is that the decay of spatial correlation is slower with distance for 
    the high enstrophy region as compared to the high dissipation regions which must be more ``broken up'' and less coherent,
    consistent with what is seen in the visualization, Fig. \ref{fig:excursion-20-vol}(b).  We conjecture that the slow correlation decay reflects the underling 
    highly elongated structure of high vorticity regions, which is not the same for most of the other observables. 
      
  \section{Joint Iso-set analysis} \label{sec:joint}
      As done for a single scalar, it is possible to define joint excursion sets for both enstrophy 
    $\omega^2/2$ and dissipation $S^2 = S_{ij} S_{ij}$, according to:
            \begin{eqnarray}
        \Theta^{\omega \epsilon}_{\chi_{\omega},\chi_{\epsilon}}({\bf x}) &=& \Theta(\omega^2/2-\chi_\omega) \Theta(S^2-\chi_\epsilon) \nonumber \\ 
                                                &=& \begin{cases}
                                                      1, & \text{if } \omega^2 \geq \chi_\omega \ {\rm and} \ S^2 \geq \chi_\epsilon\\
                                                      0, & \text{otherwise}
                                                    \end{cases}
      \end{eqnarray}
      
\begin{figure}[H]
        \begin{minipage}{0.49\linewidth}
          \includegraphics[width=\linewidth]{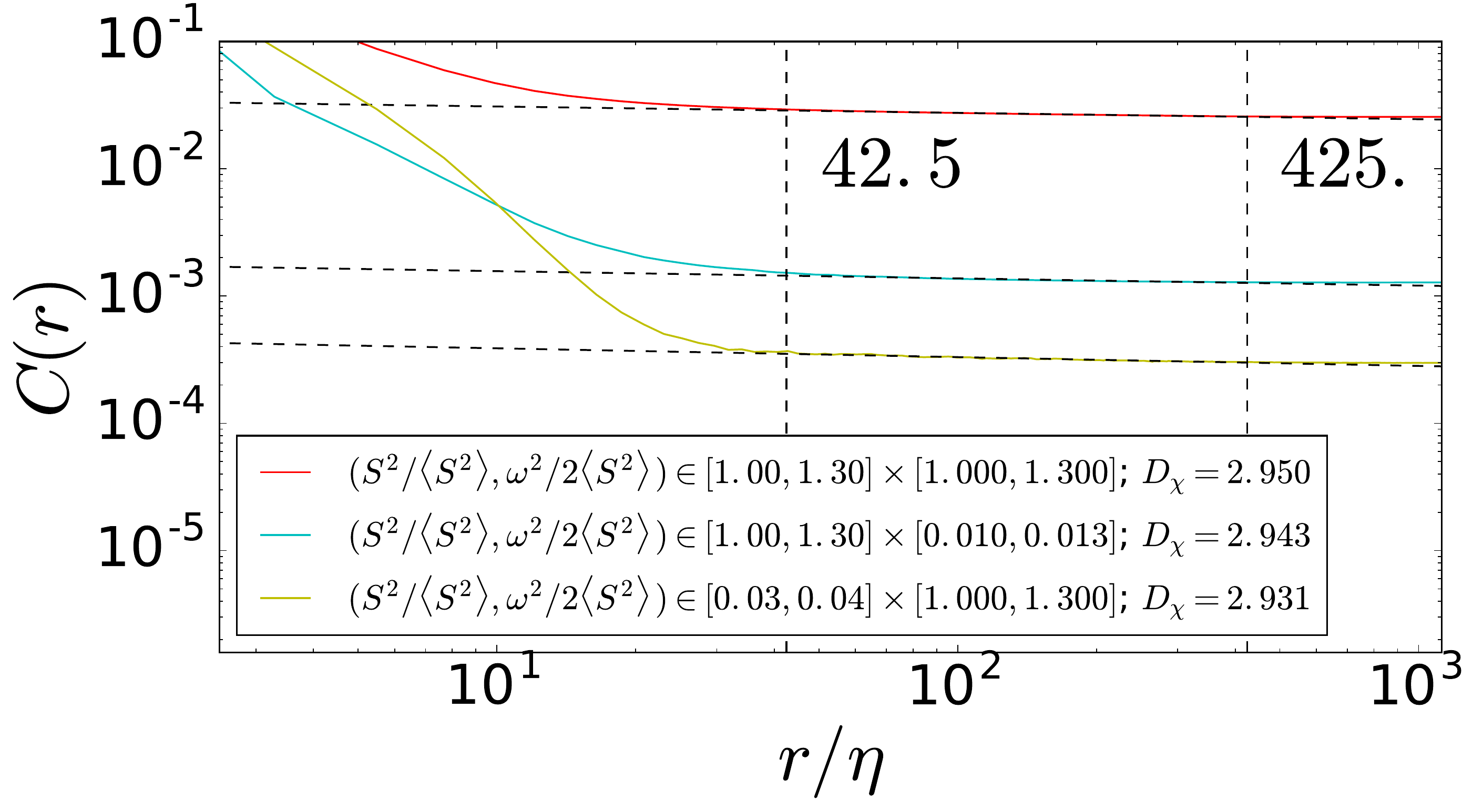}
          \put (-185,92){\makebox[0.05\linewidth][r]{\large (a)}}
        \end{minipage}
        \begin{minipage}{0.49\linewidth}        
          \includegraphics[width=\linewidth]{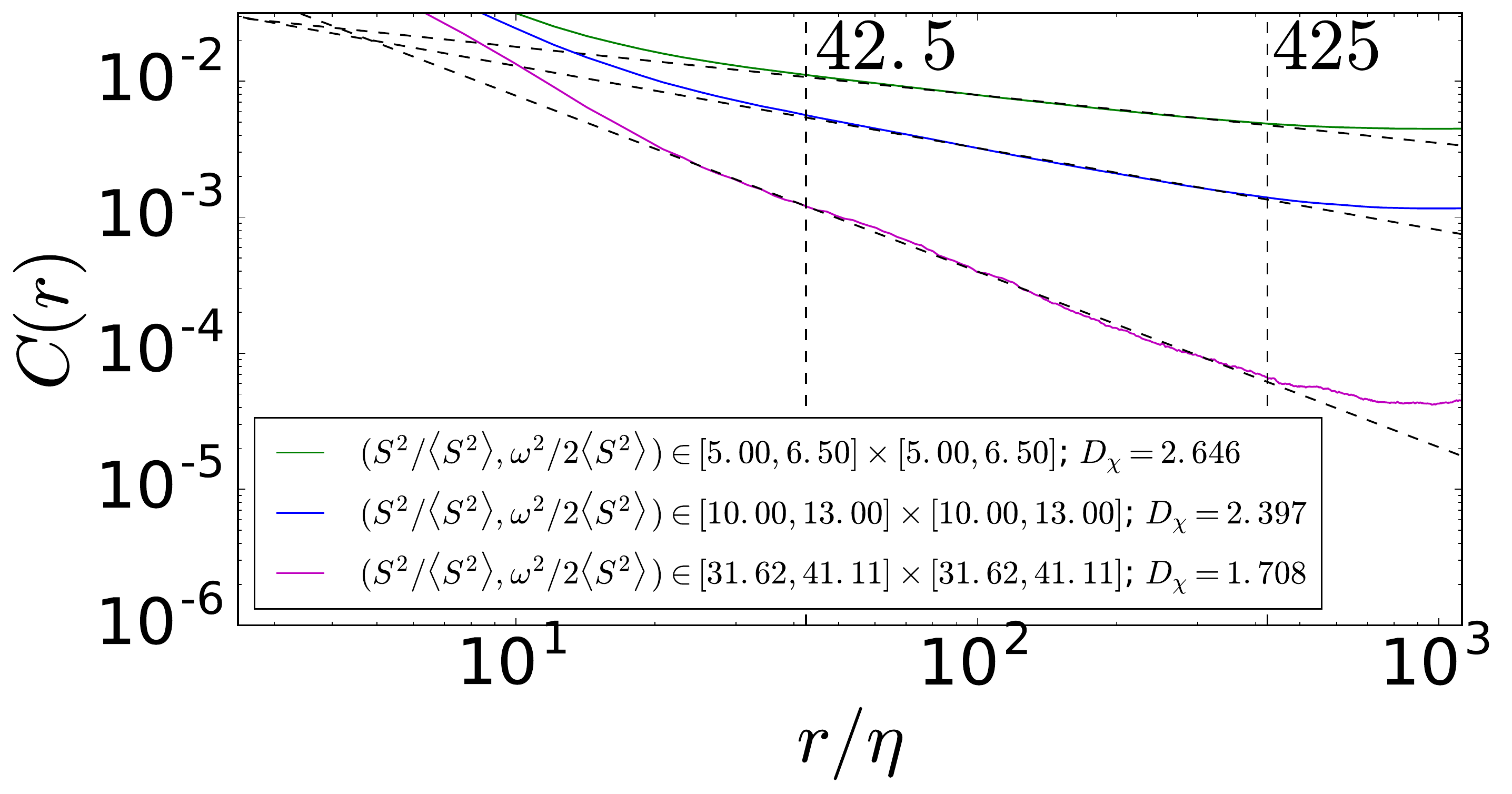}
          \put (-185,102){\makebox[0.05\linewidth][r]{\large (b)}}
        \end{minipage}
        
         \caption{Set of representative log-log plots showing power law scaling of the correlation functions for joint shell 
                  sets on enstrophy $\omega^2/2$ and dissipation $S^2$}
        \label{fig:joint-interval-samples}
      \end{figure}

      The radial correlation function of these sets are computed following the same approach as in the previous section (\S \ref{sec:excursion}). 
      Representative results are shown in figures \ref{fig:joint-interval-samples}(a,b). We observe the same overall power-law behavior 
      seen in the single excursion sets, in the same range of length-scales corresponding to the inertial range.        
      Similarly, we can define joint interval-based sets according to 
      
      \begin{eqnarray}\label{eq:joint-interval}
        \Theta^{\omega \epsilon}_{\chi_{\omega},\chi_{\epsilon},\Delta}({\bf x}) &=& \Theta(\omega^2/2-\chi_\omega)\Theta(\chi_\omega(1+\Delta_\omega)-\omega^2/2) \Theta(S^2-\chi_\epsilon) \Theta(\chi_\epsilon(1+\Delta_\epsilon)-S^2) \nonumber \\ 
                                                &=& \begin{cases}
                                                      1, & \text{if } \chi_\omega \leq \omega^2/2 \leq \chi_\omega(1+\Delta_\omega)\ {\rm and}\ \chi_\epsilon \leq S^2 \leq \chi_\epsilon(1+\Delta_\epsilon)\\
                                                      0, & \text{otherwise}
                                                    \end{cases}
      \end{eqnarray}
      
\begin{figure}[H]
        \begin{center}
          \includegraphics[width=0.6\linewidth]{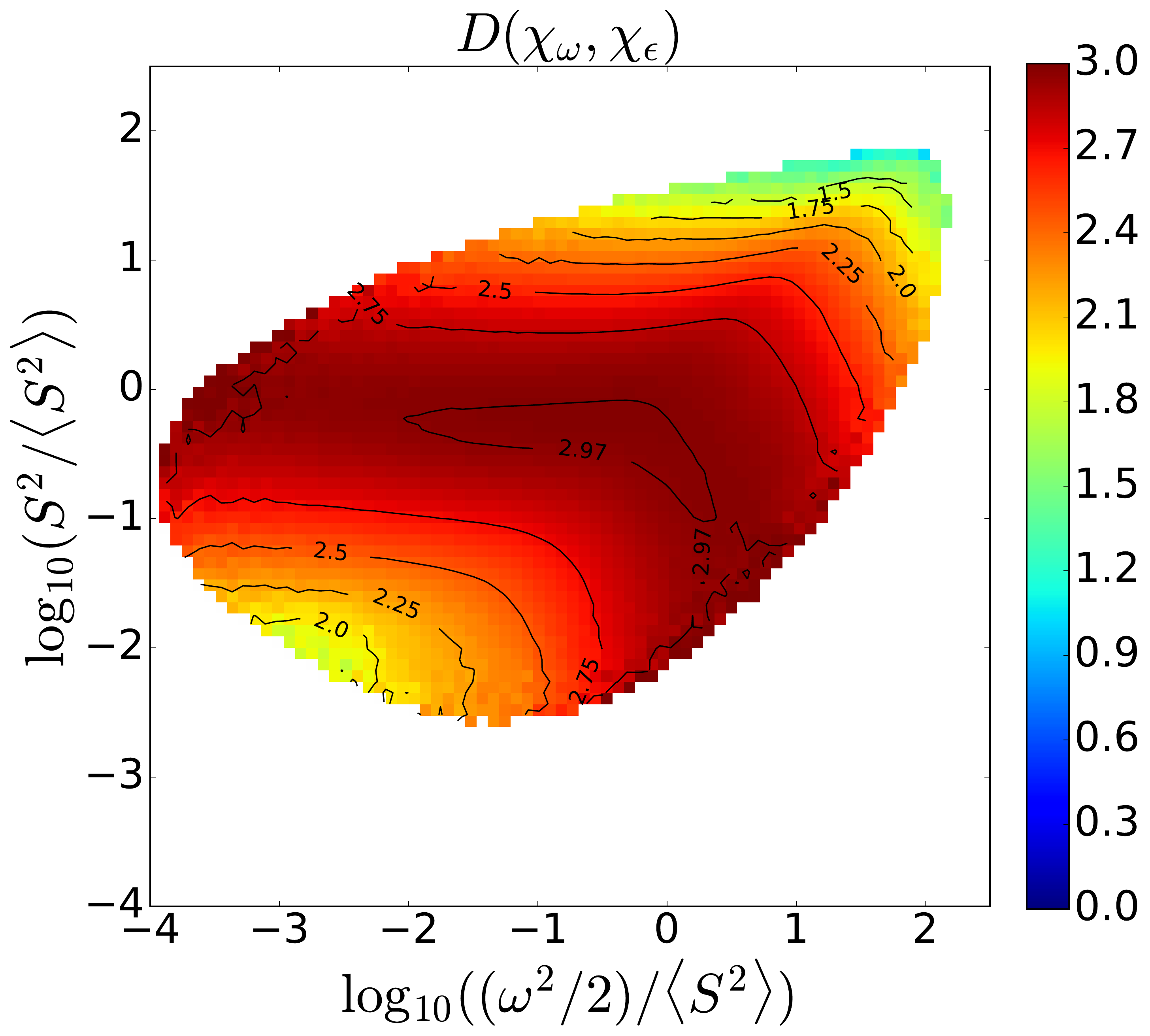}
          \caption{Joint correlation function  exponent 
                 $D(\chi_\omega,\chi_\epsilon) = 3 - \gamma_{\chi_\omega,\chi_\epsilon}$ for joint interval-based sets, with
                 equally spaced logarithmic bins $\Delta = \Delta_\omega = \Delta_\epsilon \approx 0.28$, with the same bins 
                 utilized to compute the joint-PDF in figure (\ref{fig:enst-strain-pdf}c).}       
        \label{fig:joint-interval-dims}   
        \end{center}
      \end{figure}

      For very small spacings $\Delta$, there might be numerical and statistical problems due to the very small number of 
    points on a finite dataset. Therefore, the map of the joint 2-point correlation function is only accurate for the center 
    most region, away from the skirt of the joint probability distribution function, in figure (\ref{fig:joint-interval-dims}).
    Though not shown here, we also performed sensitivity analysis to the bin size, analogously to the presented in figure
    \ref{fig:corr-enstr-strain-diff}(c,f),  and similar results were obtained. 
    
      One feature we observe in figure (\ref{fig:joint-interval-dims}) is the presence of an inverted/rotated ``L-shaped" 
    region of constant dimension, indicating a near independence of the geometrical distribution of one of the observables. 
    This indicates that the regions with either average enstrophy or dissipation are dominated by space-filling geometry, 
    irrespective of the value of the other quantity within those regions. As regions of very high or very low enstrophy and 
    dissipation are probed, we observe lower and lower correlation dimension 
    $D(\chi_\omega,\chi_\epsilon) = 3 - \gamma_{\chi_\omega,\chi_\epsilon}$, as expected. The lowest observed compensated 
    exponent on the probed region is around $D(\chi_\omega,\chi_\epsilon) \approx 1.2$ . 
      
      Considering the values along the diagonal where $S^2=\omega^2/2$, or $Q=0$, i.e. $D_{\rm diag}(\chi)= D(\chi,\chi)$, 
    we can approximate the joint distribution only in terms of this function, i.e. 
    $D(\chi_\omega,\chi_\epsilon) \approx D_{\rm diag}(\max\{\chi_\omega,\chi_\epsilon\})$. 
    This approximation reproduces the L-shaped pattern quite well (not shown), suggesting that any  $Q\neq 0$
    regions have, to a first approximation, the dimension associated with the $Q=0$ hull for the component with the highest 
    intensity, of either $S^2$ or $\omega^2/2$.
      

\section{Scaling analysis of spatial distribution of  invariants $Q$ and $R$:}
      
\begin{figure}[H]
        \begin{center}

        \begin{minipage}{0.32\linewidth}
          \includegraphics[width=\linewidth]{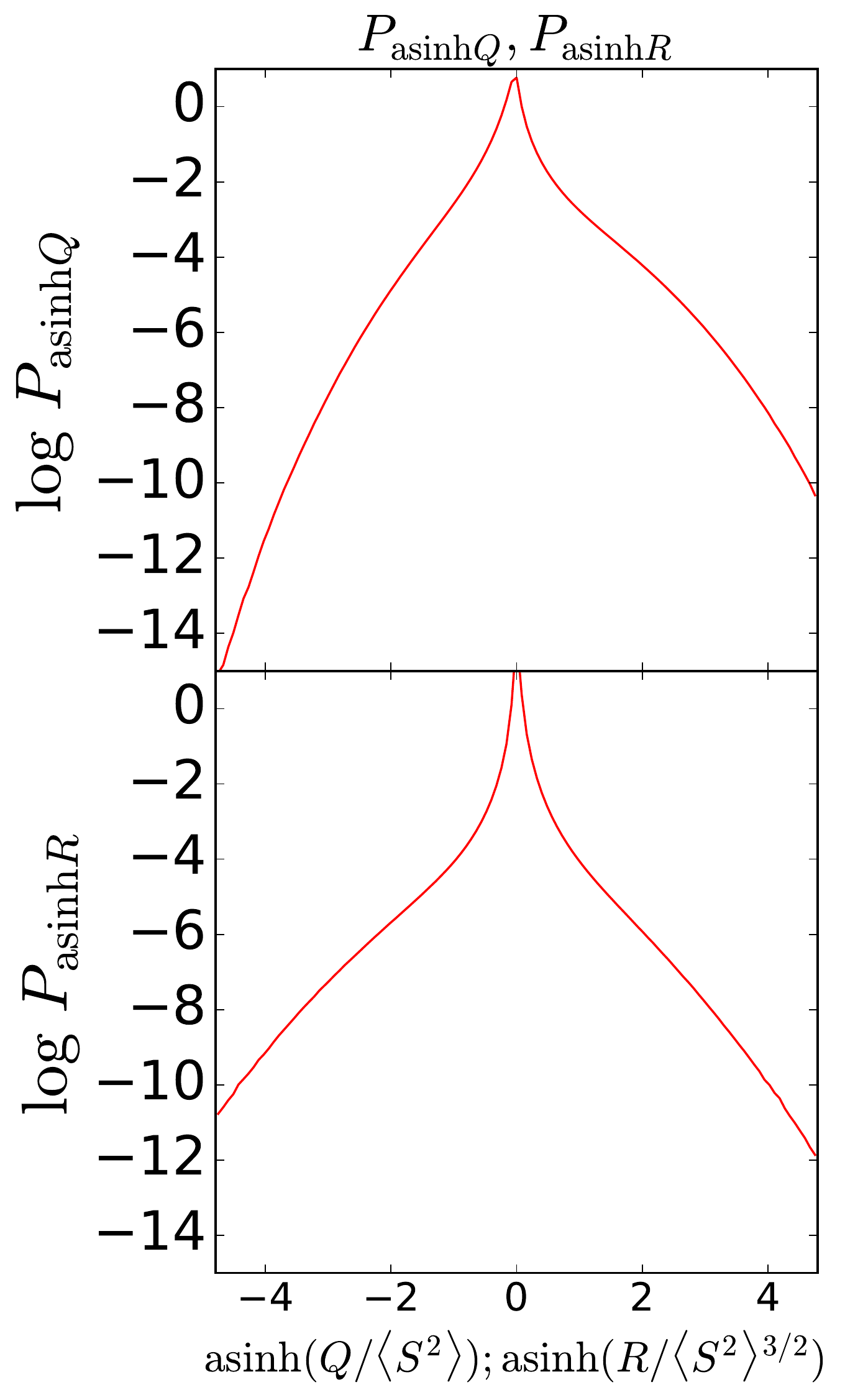}
          \put (-19,223){\makebox[0.05\linewidth][r]{(a)}}
          \put (-19,116){\makebox[0.05\linewidth][r]{(b)}}
        \end{minipage}
        \begin{minipage}{0.07\linewidth}
        \end{minipage}
        \begin{minipage}{0.60\linewidth}
          \includegraphics[width=\linewidth]{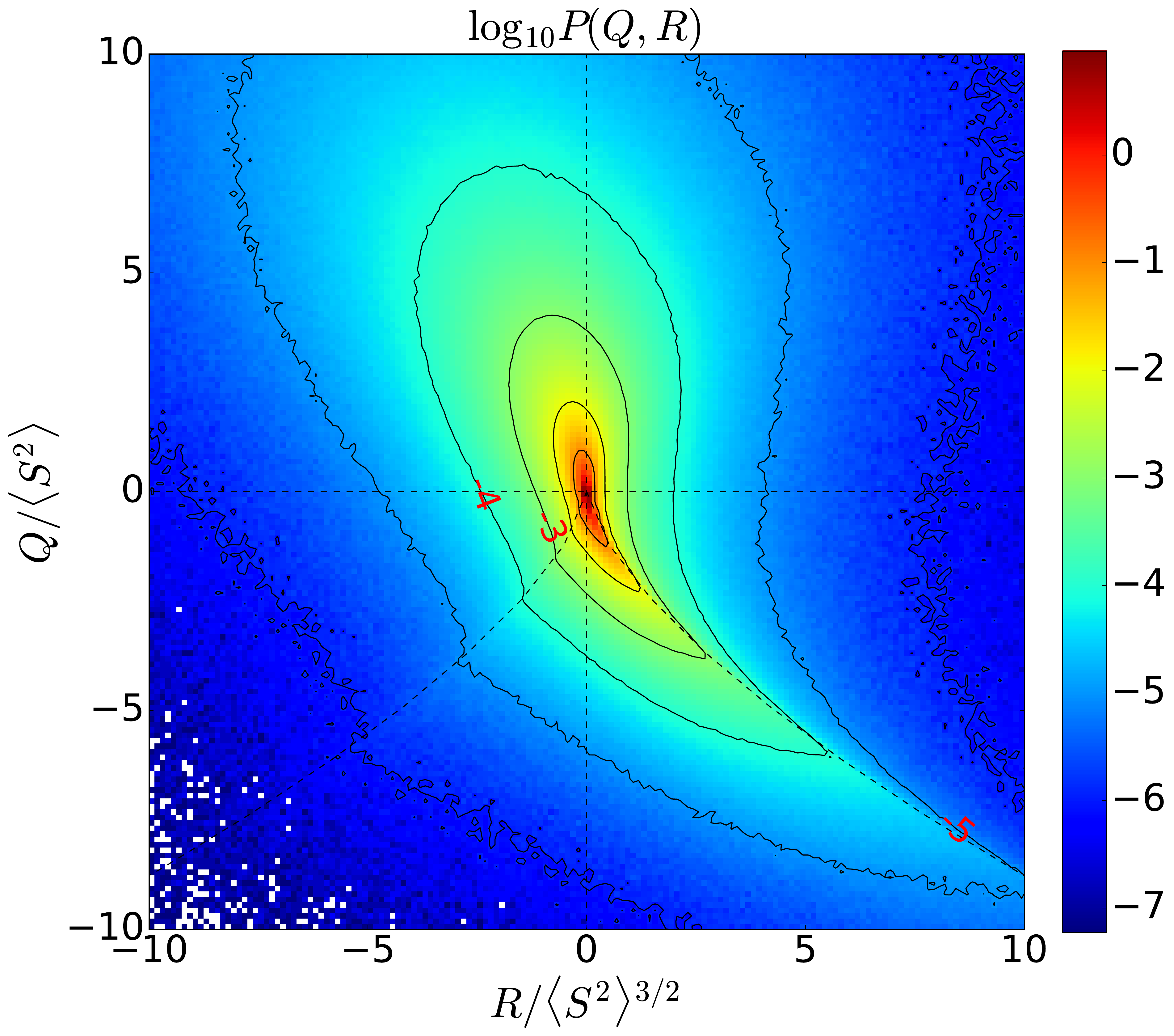}
          \put ( -56,222){\makebox[0.05\linewidth][r]{\Large \color{white}{(c)}}}
        \end{minipage}
         
        \end{center}
          
        \caption{PDFs of $Q$ (a) and $R$ (b) PDFs. (c) $Q$ and $R$ Joint-PDF, a similar joint PDF can be seen in 
                  reference \cite{meneveau2011lagrangian}.}
        \label{fig:qr-pdf}
      \end{figure}

      Following the work done in the previous sections, we seek to probe the geometrical structure of the observables $Q$ and
    $R$ defined in equation \eqref{eq:QR-def}. Since both quantities are signed, we computed their PDFs as function of 
    thresholds in linear instead of logarithmic scale. First, we present the  PDFs of both quantities, in Fig. 
    \ref{fig:qr-pdf}(a,b), and   the joint PDF of $Q$ and $R$ in Fig \ref{fig:qr-pdf}(c). We notice the characteristic 
    tear-drop shape in the joint PDF, with the right-most region following the so called Vieillefosse tail as 
    $Q = -\frac{3}{2^{3/2}} R^{2/3}$. More details can be found in Refs.
    \cite{vieillefosse1982local,meneveau2011lagrangian,pumir2007,ooi1998dynamics}.
    
      The quantities in Fig. \ref{fig:qr-pdf}(a,b) where ploted on inverse hyperbolic sine (${\rm asinh}$) axis. The asymptotic 
    behavior of ${\rm asinh}$ for large values is to approach $\log$, while being linear close to the origin. These features 
    allow us to have a reasonably undistorted view of the PDF near the origin, and also verify if there is any power law 
    behavior on the tails of the PDF, for either positive or negative values of the quantities of interest. In this case it
    appears that no power-law behavior is visible in the tails of the PDFs of $Q$ and $R$, on either positive or negative 
    sides.
            
\begin{figure} [H]
              \begin{center}
          \begin{minipage}{0.4\linewidth}
          \includegraphics[width=\linewidth]{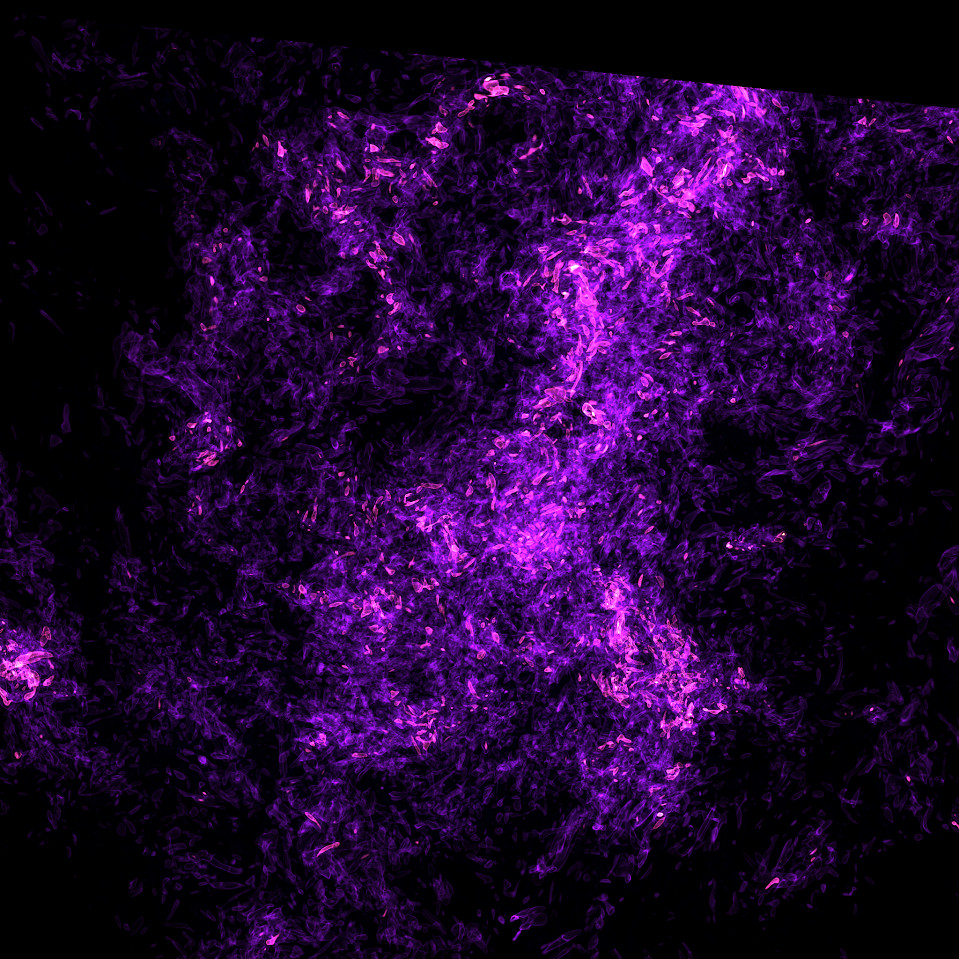}
           \put (-16,216){\makebox[0.05\linewidth][r]{\Large \color{white}{(a)}}}  
         \end{minipage}
         \begin{minipage}{0.4\linewidth}
         \includegraphics[width=\linewidth]{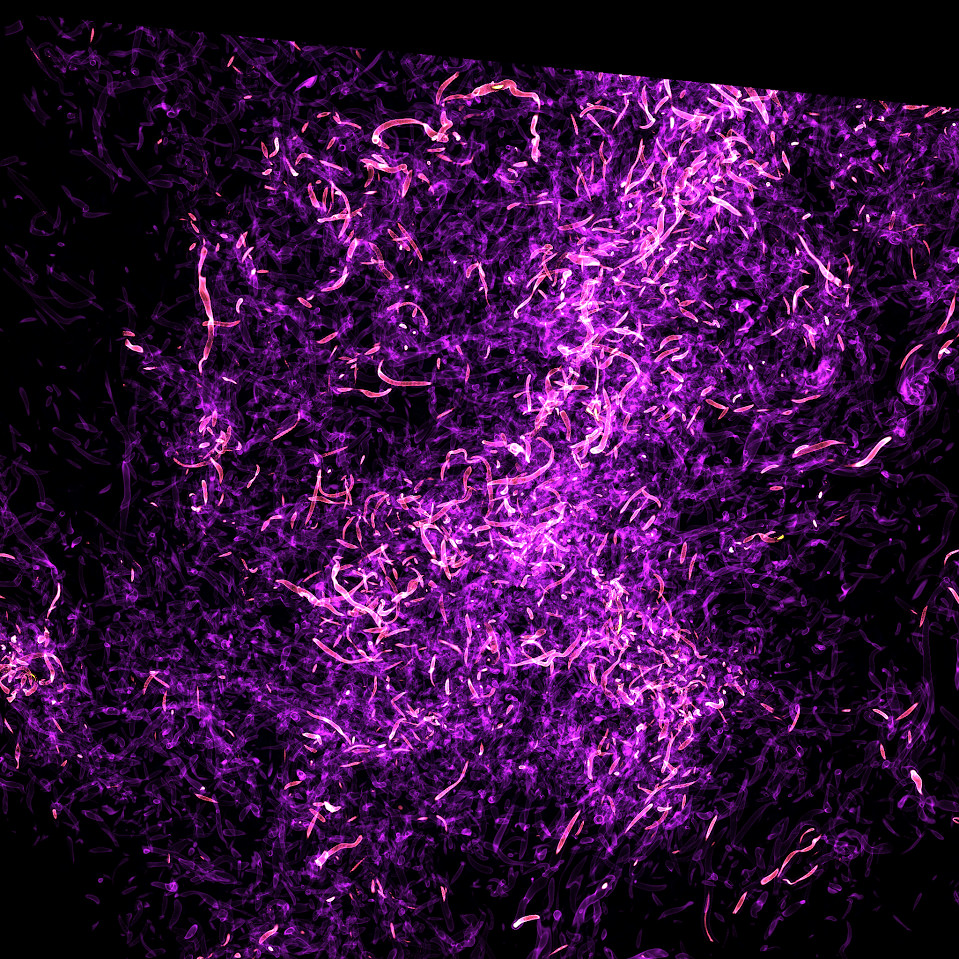}
          \put (-16,216){\makebox[0.05\linewidth][r]{\Large \color{white}{(b)}}}     
         \end{minipage}
               \end{center}
          \caption{Volume rendering of (a) the $Q$ velocity gradient invariant excursion set for negative $Q<-2\langle S^2\rangle$, i.e. corresponding to the function \\
                  $Q_{ex}({\bf x}) = -Q({\bf x}) \Theta_{2\langle S^2\rangle}^{-Q}({\bf x})$ 
                  and (b) the $Q$  excursion set for positive $Q>2\langle S^2\rangle$, i.e. corresponding to   
                  the function $Q_{ex}({\bf x}) = Q({\bf x}) \Theta_{2\langle S^2\rangle}^Q({\bf x})$
                  on a $512^3$ subset of the full data.}        
       \label{fig:q-volume rendering}
       \end{figure}

      Before proceeding to analyze the spatial correlation functions of the corresponding shell sets, it is useful 
    to present  visualizations of the $Q$ and $R$ scalar fields. In figure \ref{fig:q-volume rendering} we observe 
    that the overall, middle and large scale spatial distribution strongly resembles the ones present earlier in 
    figure \ref{fig:excursion-20-vol}, especially comparing Fig. \ref{fig:excursion-20-vol}(a) and Fig. 
    \ref{fig:q-volume rendering}(b). This resemblance is expected since $Q>0$ thresholds are often used as vortex 
    visualizations (the Q-criterion \cite{hunt1988eddies}).   Negative $Q$ regions are more correlated with high 
    straining region, again as expected based on the identity  $Q=\omega^2/2 - S^2$. 
    
      \begin{figure} [H]
            \begin{center}
        \begin{minipage}{0.4\linewidth}
        \includegraphics[width=\linewidth]{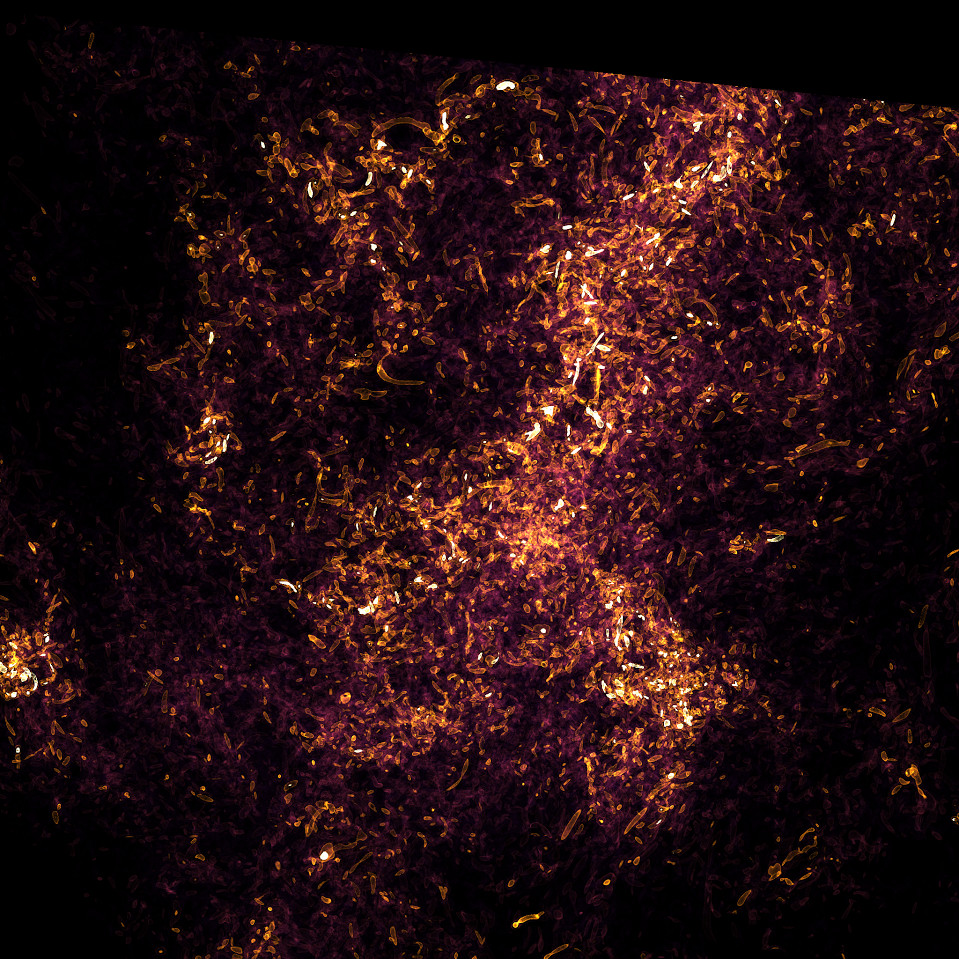}
          \put (-16,216){\makebox[0.05\linewidth][r]{\Large \color{white}{(a)}}}     
        \end{minipage}
        \begin{minipage}{0.4\linewidth}
        \includegraphics[width=\linewidth]{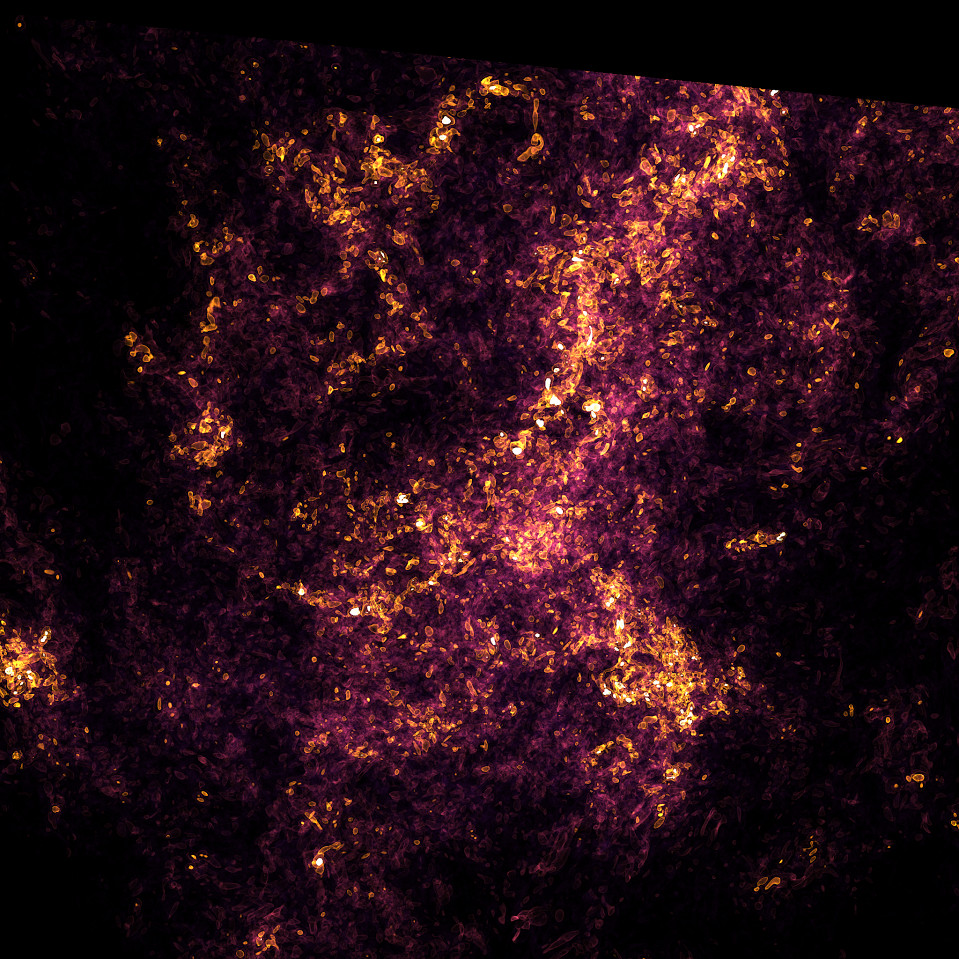}
          \put (-16,216){\makebox[0.05\linewidth][r]{\Large \color{white}{(b)}}}     
        \end{minipage}
              \end{center}
         \caption{Volume rendering of (a) the $R$ velocity gradient invariant excursion set corresponding to the function \\
                 $R_{ex}({\bf x}) = -R({\bf x}) \Theta_{2\langle S^2\rangle}^{-R}({\bf x})$ 
                 and (b) of $R$ set corresponding to  
                 the function $R_{ex}({\bf x}) = R({\bf x}) \Theta_{2\langle S^2\rangle}^R({\bf x})$
                on a subset of the full cube, with $1/8$ in volume. }
        \label{fig:r-volume rendering}
      \end{figure} 
      
      Visualizations of spatial distributions of the scalar $R$ are less common in the literature (although see discussion in 
    Ref.  \cite{perryChongCantwell}). 
    Interestingly, we observe that negative $R$, in figure \ref{fig:r-volume rendering}(a), include
    slightly more elongated structures than the positive $R$ distributions, Fig. \ref{fig:r-volume rendering}(b). Hence, the 
    regions in which both $Q$ and $R$ show elongated structures are in the upper-left quadrant of the $RQ$ plane, the vortex 
    stretching quadrant.    
            
\begin{figure}[H]
        \begin{minipage}{0.35\linewidth}
          \includegraphics[width=\linewidth]{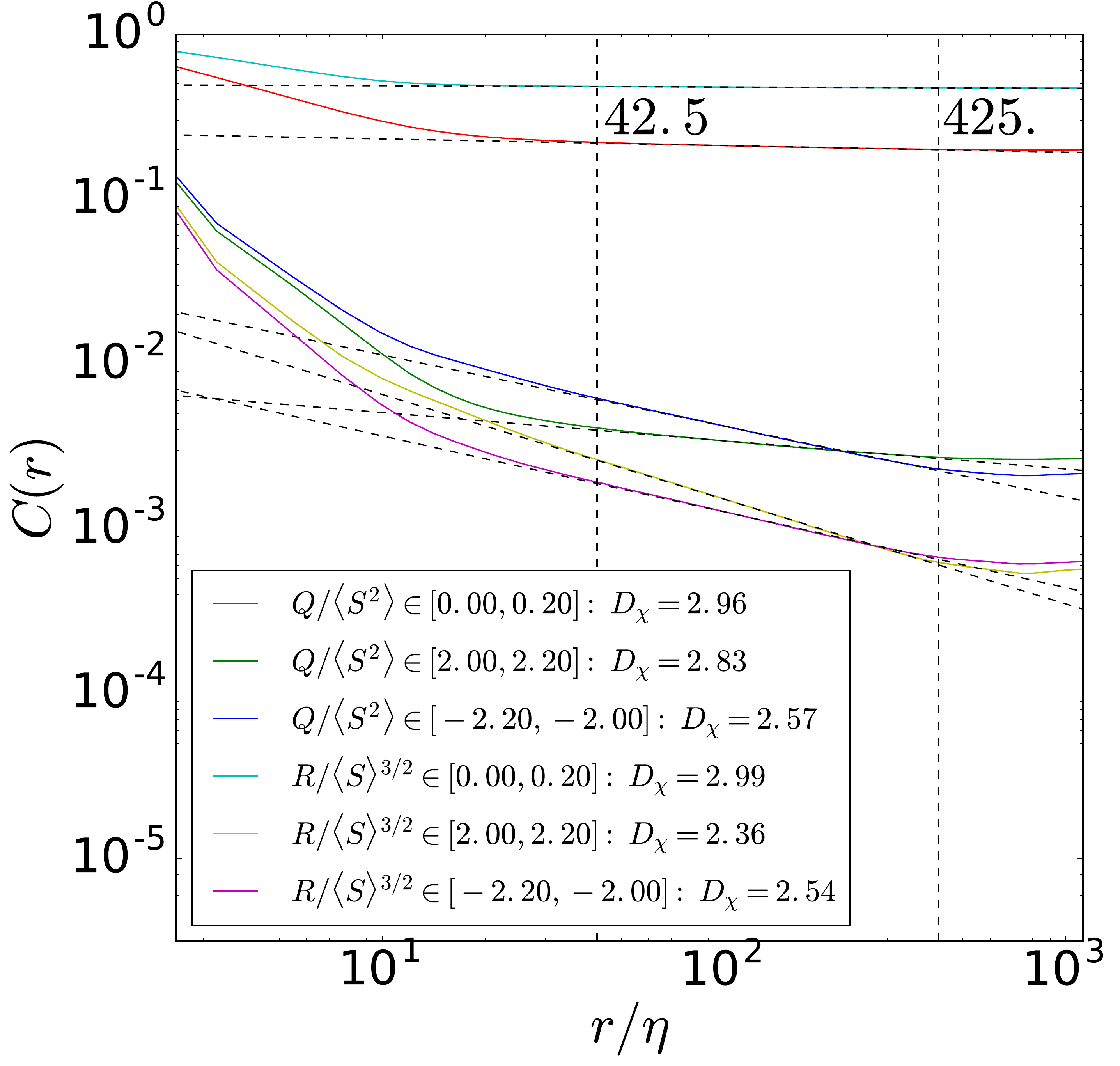}
          \put (-15,115){\makebox[0.05\linewidth][r]{\Large (a)}}
        \end{minipage}
        \begin{minipage}{0.64\linewidth}
          \includegraphics[width=\linewidth]{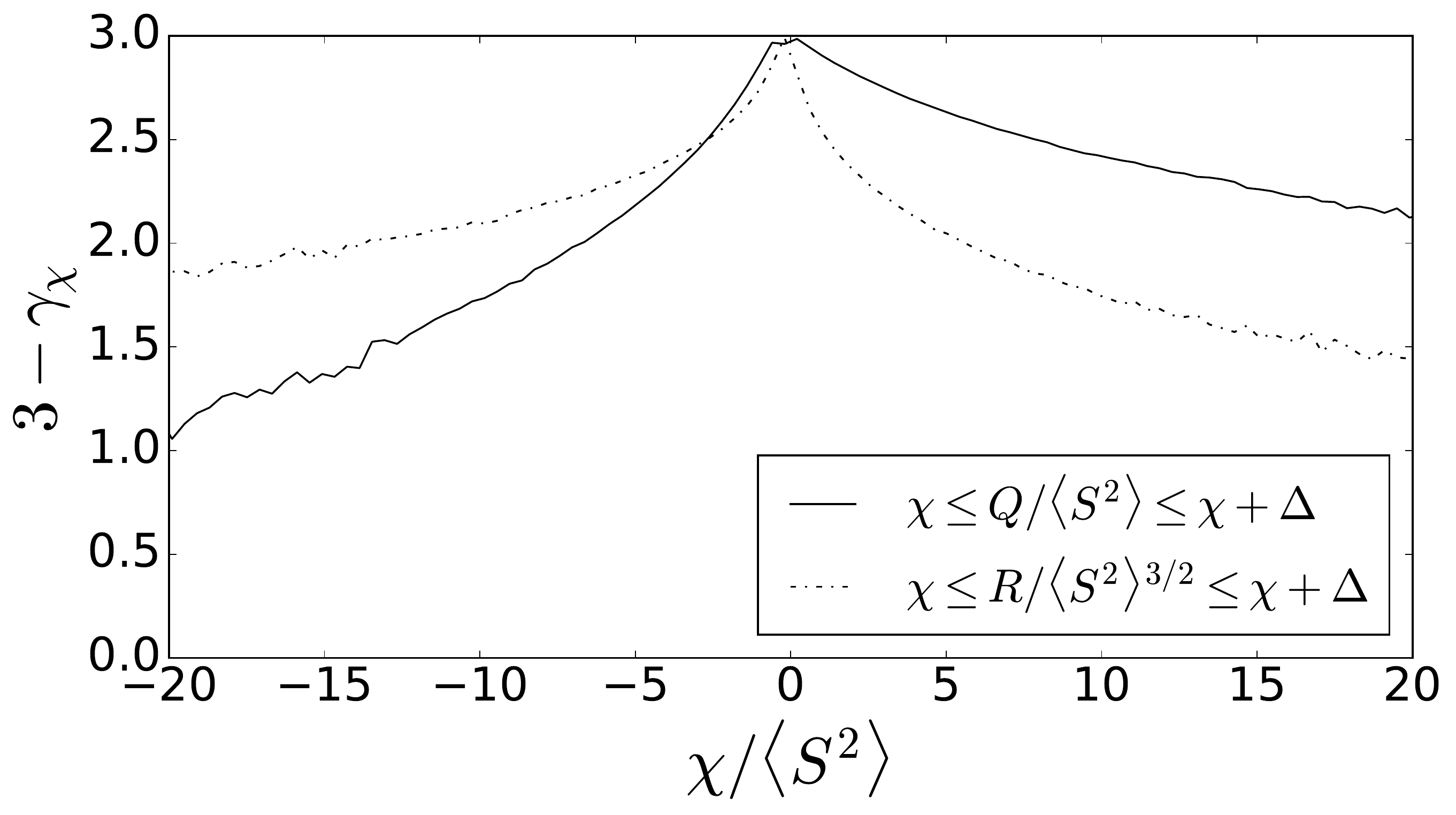}
          \put (-27,147){\makebox[0.05\linewidth][r]{\Large (b)}}
        \end{minipage}        
      
        \caption{(a):  Correlation functions for selected values of $Q$ and $R$. 
                 (b): Correlation-based dimensions of joint interval-based sets, as a function of thresholds on
                 $Q$ and $R$ .}
        \label{fig:qr-interval-dims}
      \end{figure}

      To quantify the spatial correlation structure, the correlation functions of interval-sets are computed as before, for 
    various thresholds of $Q$ and $R$. Similarly to what is observed for enstrophy and dissipation, we find clear
    power-laws in the two-point correlation functions associated with the iso-sets of $Q$ and $R$, as exemplified in figure
    \ref{fig:qr-interval-dims}(a).  
    
    The measured correlation dimensions, as a function of the threshold 
    $\chi$, are presented in figure Fig. (\ref{fig:qr-interval-dims}b). The basic behavior of the 
    correlation dimension mimics the PDF of the corresponding observable, as can be seen comparing 
    \ref{fig:qr-pdf}(a,b) and \ref{fig:qr-interval-dims}(b).

\begin{figure}[H]
        \begin{minipage}{0.49\linewidth}
          \includegraphics[width=0.9\linewidth]{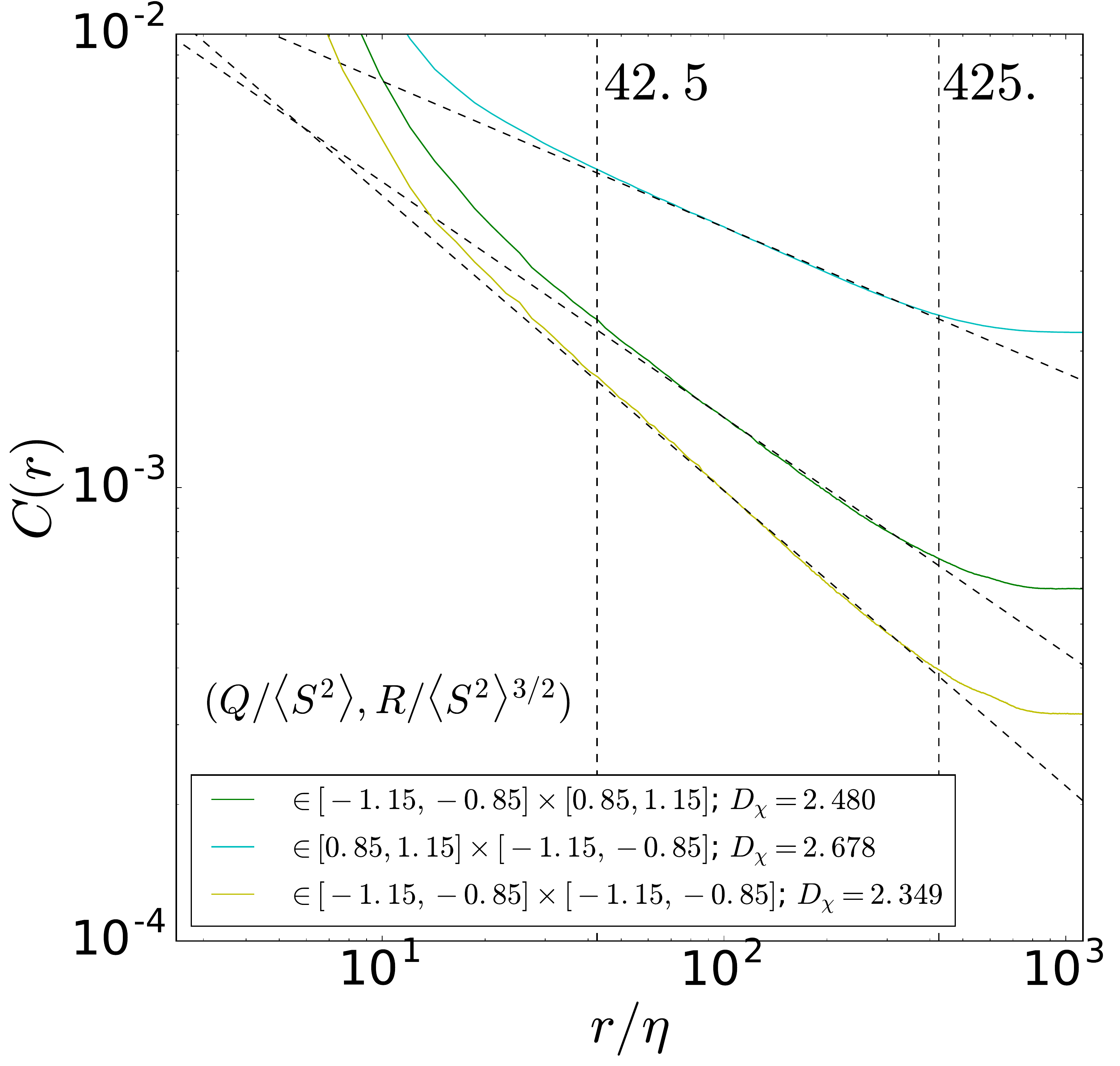}
          \put (-20,110){\makebox[0.05\linewidth][r]{\Large (a)}}
        \end{minipage}
        \begin{minipage}{0.49\linewidth}        
          \includegraphics[width=0.9\linewidth]{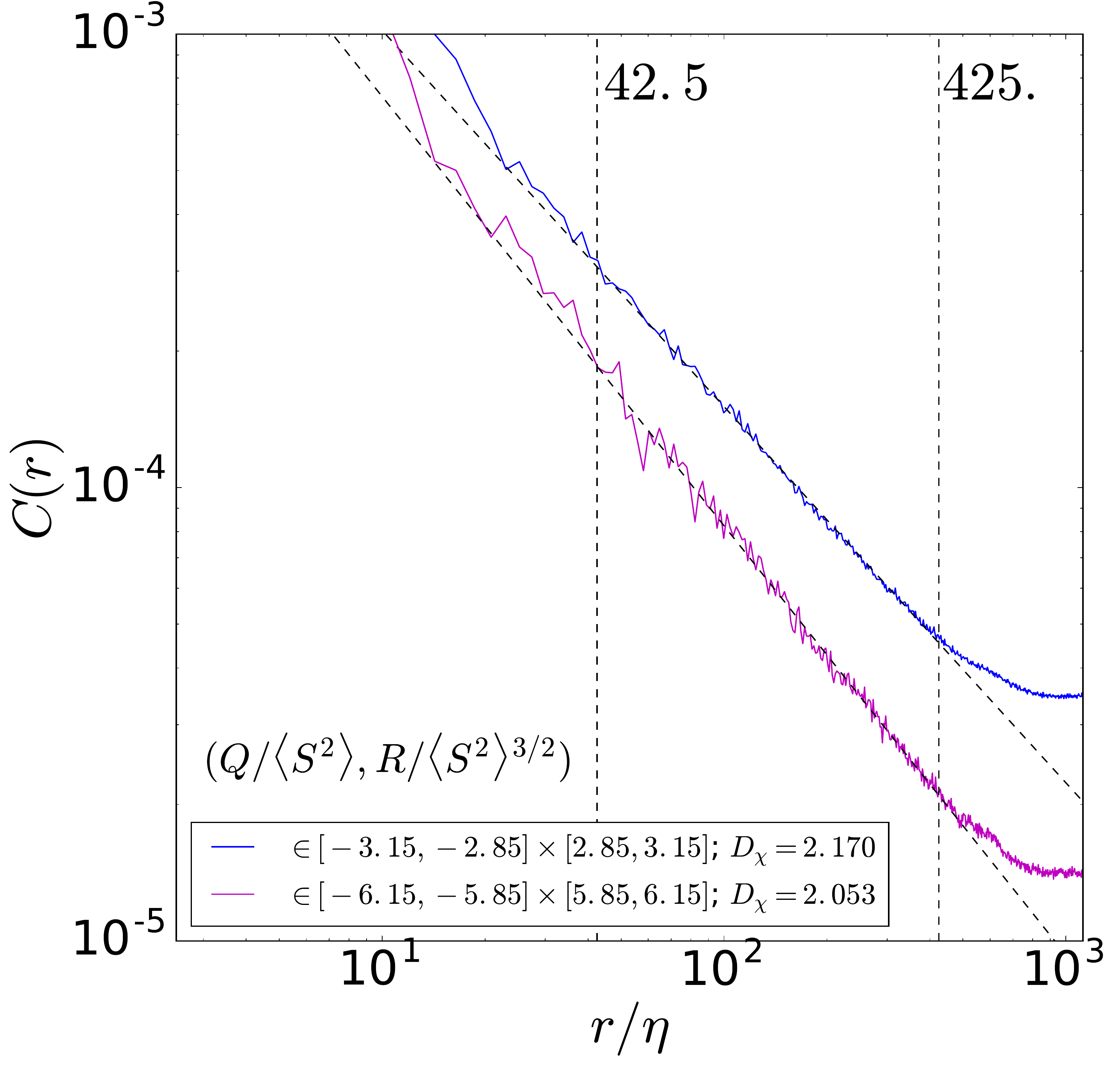}
          \put (-20,110){\makebox[0.05\linewidth][r]{\Large (b)}}
        \end{minipage}
      
         \caption{Set of representative log-log plots for power law of the correlation functions for joint shell sets 
                  for $Q$ and $R$ velocity gradient invariants}
        \label{fig:qr-joint-interval-samples}   
      \end{figure}

\begin{figure}[H]
        \begin{center}
          \includegraphics[width=0.7\linewidth]{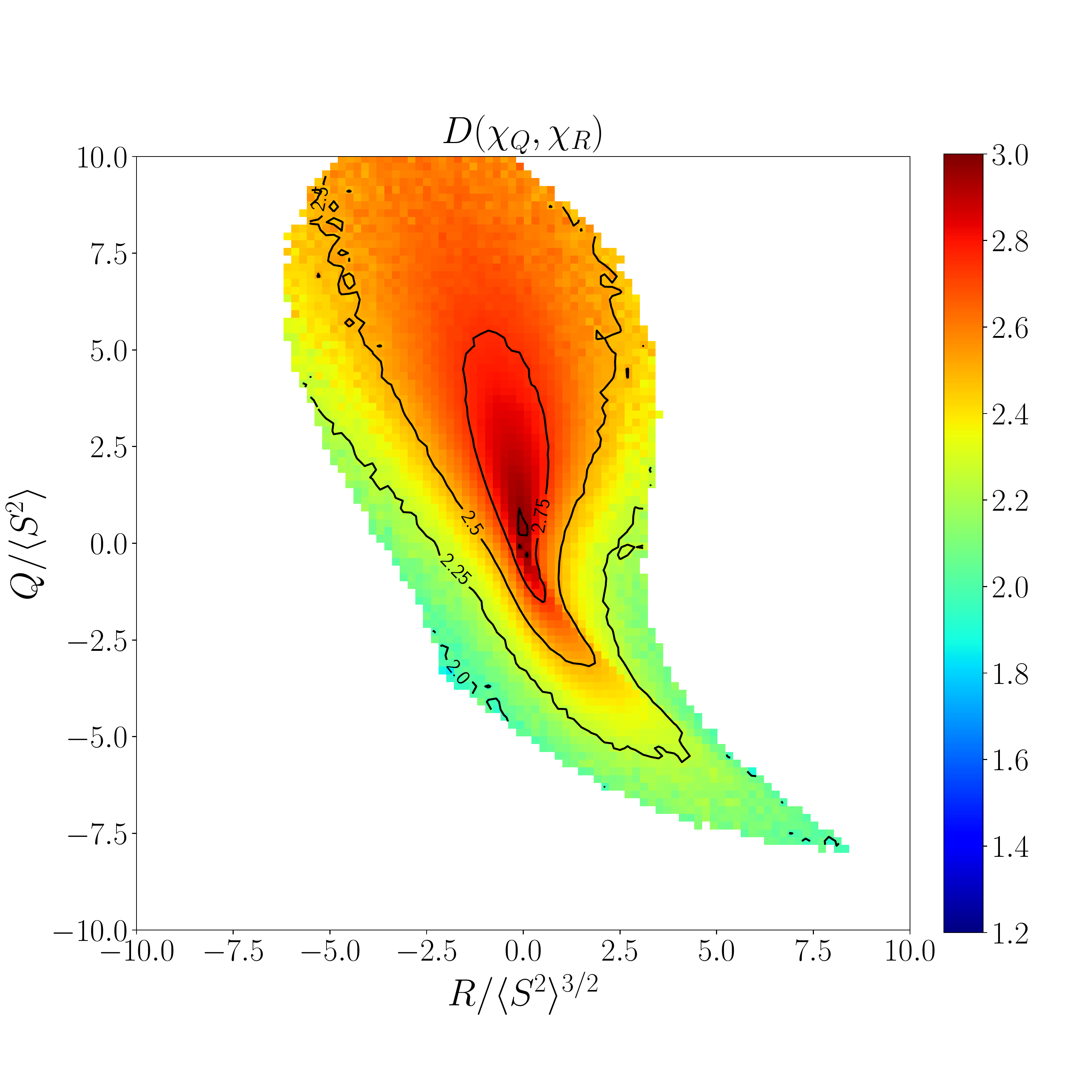}
          \put (-83,262){\makebox[0.05\linewidth][r]{\Large {\bf (c)}}}     
          
          \vspace*{-3em}
        
          \caption{Joint correlation function  exponent 
                 $D(\chi_\omega,\chi_\epsilon) = 3 - \gamma_{\chi_Q,\chi_R}$ for joint interval-based sets, with
                 equally spaced linear bins $\Delta = \Delta_Q = \Delta_R = 0.2$.}
        \label{fig:qr-joint-interval-dims}   
        \end{center}
      \end{figure}

        For the analysis of joint $Q$ and $R$ sets, we present some representative log-log plots in Fig. 
    \ref{fig:qr-joint-interval-samples}(a,b), which showcase that the correlation function presents power-law behavior for 
    these sets as well. The full joint correlation dimension distribution is presented in Fig. 
    \ref{fig:qr-joint-interval-dims}.
        
      \vspace*{-3em}        
      
      The most striking feature of Fig. (\ref{fig:qr-joint-interval-dims})  is the top-bottom asymmetry of the $Q>0$ and $Q<0$ 
    regions for the correlation-based dimension. The dimension is clearly larger in the rotation dominated regions at $Q>0$.  
    This is consistent with the fact that, for the same threshold, correlation dimensions associated with enstrophy iso-sets 
    are higher than dissipation iso-sets (Fig.  \ref{fig:interval-dims}(b)). Clearly the geometric features of the joint 
    distribution differs from the joint PDF in Fig. \ref{fig:qr-pdf}(c).  That is to say, regions with high PDF need not 
    have higher (more space-filling) correlation-based dimension. 
          
  \section{Conclusions} \label{sec:conclusion}
    
     We have analyzed a turbulence dataset from DNS at a moderately high Reynolds number, with the specific aim to 
   identify scaling laws characterizing the spatial distribution of phenomena of various magnitudes. Both excursion sets and 
   iso-sets (thin bands) have been considered for the enstrophy, dissipation (or square-strain-rate), and the two invariants 
   $Q$ and $R$. The variable $Q$ has also often been used for flow visualization and high positive values of $Q$ can be used 
   to identify vortices. The spatial distributions are first defined using an indicator function and the radial correlation 
   function of the indicator function distribution is evaluated. In all cases we find clear power-law decay in these 
   correlations for separation distances falling within the inertial range of turbulence. The scaling range is insensitive to 
   the thresholds and variable of interest. We confirmed this is the same scaling range characterizing power-law scaling of 
   the traditional velocity structure function. 
     
     Even though the correlation functions present power-law in the inertial range, no such power-law behavior is observed
   in box-counting and box-counting based correlation dimension computations. Thus the interpretation of the  
   correlation-function based exponents $D(\chi)=3-\gamma$ as a ``dimension'' must be considered with care. Consistent 
   with the definition of a dimension, for thresholds near the mean value, space fillingness is observed with the exponent 
   saturating at 3. For higher (or lower) thresholds, this correlation dimension reduces to lower values.  
    
     We also observe some surprising trends, such as a lower correlation-function based dimension for strong dissipation 
   events compared to strong enstrophy events. It is likely that this is caused by the elongated nature of vortices causing 
   coherence in space over longer distances on average as compared to regions of high dissipation.   We also show that sets 
   defined by joint conditions on strain and enstrophy, and on $Q$ and $R$,  also display power law scaling in the correlation 
   functions, providing  further characterization of the complex spatial structure of the intersections of these sets.
   
   The inertial range power-law behavior of correlation functions associated with quantities in the viscous range (dissipation, 
   enstrophy, $Q$ and $R$) of a wide range of thresholds provides further evidence of geometric self similarity of flow 
   properties in the inertial range 
   
    Overall, this work shows an alternate route to study multifractal behavior of turbulence, in which geometrical information 
   is probed explicitly by using correlation functions of indicator functions. It is not yet immediately clear how to naturally
   connect the results of the present work with the traditional multi-fractal formalism, which is based on the scaling of
   statistical high-order moments of the box-averaged flow quantities over regions of different sizes. Specifically, it is not 
   clear how to associate the threshold $\chi$  to the parameters $\alpha$ or $h$ used in the multifractal formalism. 
   
   Further followup work should develop such correspondences, as well as examine the scaling for different (higher) Reynolds numbers. Also, extensions to non-isotropic shear flows, in which the correlation 
   functions may decay differently in different directions, would be of interest. 
           
     \subsection*{Acknowledgements}
           
       The authors are grateful to the Turbulence Research Group members for discussions and help with this project, 
     Dr. Gerard Lemson and Dr. Stephen Hamilton for their help with the SciServer system. 
     Jos\' e Hugo Elsas is grateful to the Rio de Janeiro state science funding agency FAPERJ 
     program for international Ph.D. exchange, grant number E-26/200.076/2016 and to Dr. L. Moriconi for authorizing the 
     international exchange. Alexander Szalay and Charles Meneveau are supported by NSF's CDS\&E: CBET-1507469 and 
     BigData:OCE-1633124 projects. The SciServer project is supported by NSF's DIBBS program (OAC-1261715). 
     
       SciServer is a collaborative research environment for large-scale data-driven science. It is being developed at, and 
     administered by, the Institute for Data Intensive Engineering and Science at Johns Hopkins University. SciServer is 
     funded by the National Science Foundation Award ACI-1261715. For more information about SciServer, please visit 
     http://www.sciserver.org.
           
  \bibliography{refs}

   \pagebreak
  
  \section*{Appendix A:  Analysis environment on SciServer, and notebooks}
    
    The data used in this paper is obtained from the Johns Hopkins Turbulence Databases (JHTDB). Most prior uses of JHTDB focussed on analysis of spatially localized regions, 
    for which local operations such as interpolation or finite-difference based differentiations could be done on the database system itself and delivering small amounts of data to users. In the present 
    work we desired instead to use FFTs for the analysis in order to enable us spectral accuracy for derivative evaluations, as well as efficient evaluation of the 3D correlation functions. 
    However, FFTs require access to the entire $1024^3$ fields, for which the usual access modes of JHTDB are not well suited. Instead,  
    the analysis  presented in this paper was performed on the Sciserver  cloud 
    environment, hosted by the Institute for Data Intensive Science at Johns Hopkins University (\url{http://www.sciserver.org}). 
    The goal of Sciserver is to provide a local environment for data driven science.
    Sciserver provides a 10 Gigabit Ethernet connection to the Johns Hopkins 
    Turbulence Database \cite{Lietal2008,Johnsonetal16}, which is a valuable asset to the present work since it allows to 
    easily download entire snapshots from the database.
     Sciserver was initially developed to be used in conjunction with
    the Sloan Digital Sky Survey, in the form of Skyserver, as a nearline analysis tool to the Astronomy database. It has 
    since then expanded to other areas of scientific research including Turbulence, Genomics and Oceanography. 
           
      We utilized the Compute module of Sciserver, which provides a Jupyter notebook environment running over Docker 
    containers that provide user package customizability through Anaconda and Pip package managers. Sciserver also 
    provides a set of pre-configured docker containers for Python, Matlab and other languages.  
        
      \begin{figure}[H]
        \begin{minipage}{0.48\linewidth}
          \fbox{\includegraphics[width=0.97\linewidth]{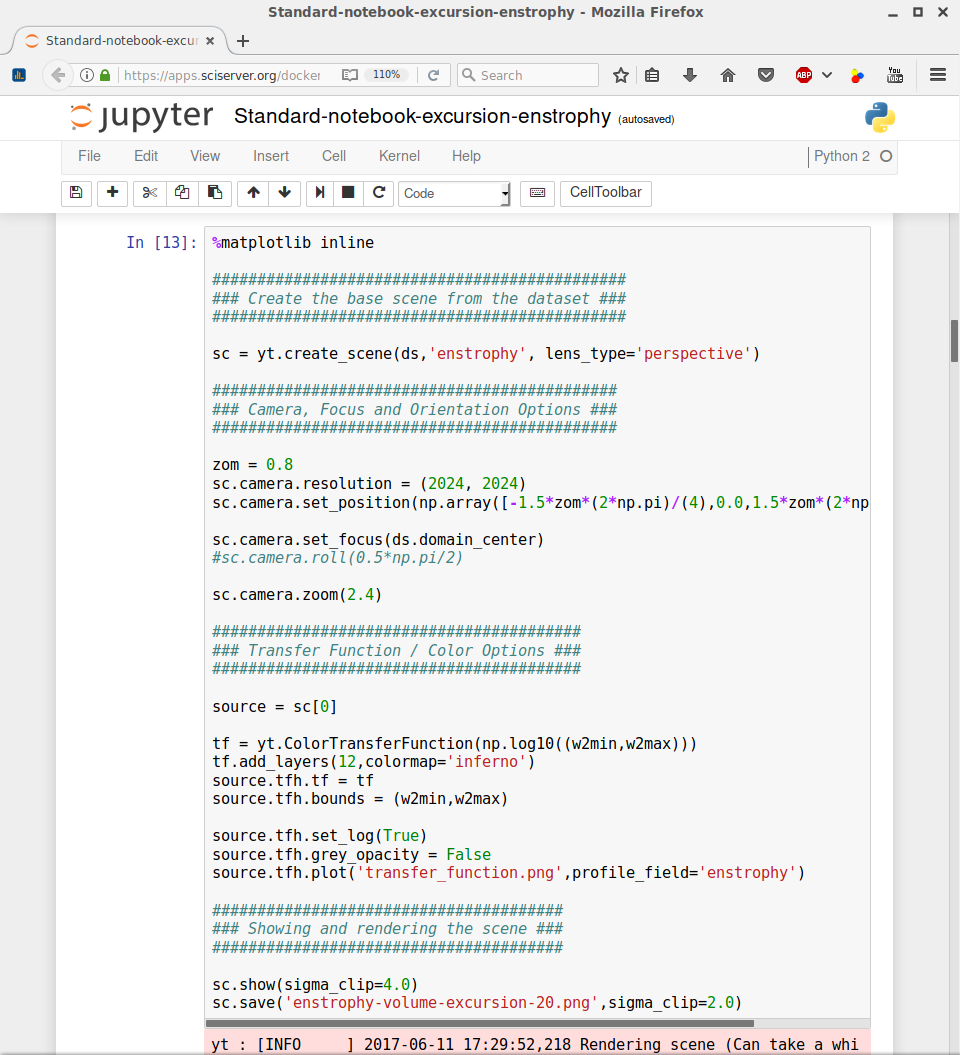}}
          \put (-13,180){\makebox[0.05\linewidth][r]{\Large {\bf \color{black}{(a)}}}}
        \end{minipage}
        \begin{minipage}{0.48\linewidth}
          \fbox{\includegraphics[width=0.97\linewidth]{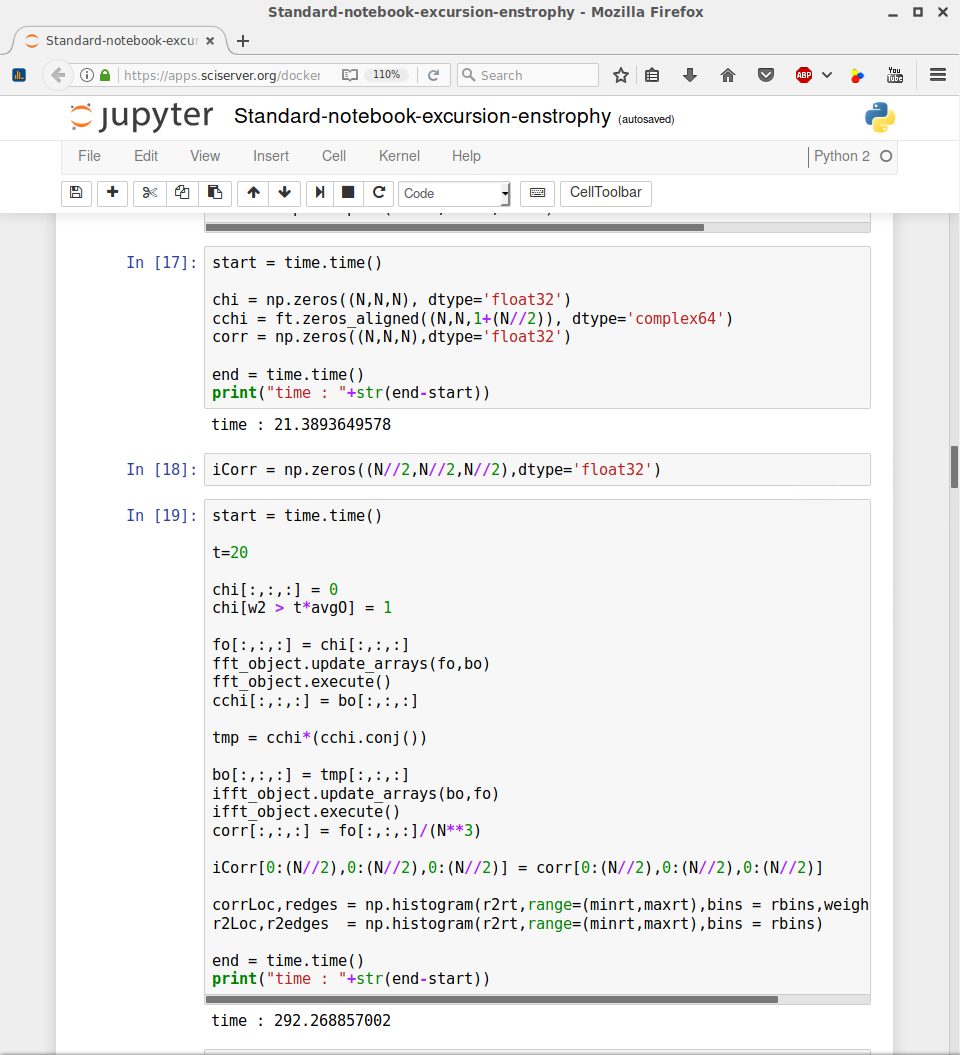}}
          \put (-13,180){\makebox[0.05\linewidth][r]{\Large {\bf \color{black}{(b)}}}}
        \end{minipage}
              
        \caption{Code snippets used in the Sciserver environment : (a) Volume rendering script with YT library; 
                                                          (b) Two-point correlation function}
        \label{fig:sciserver-1}   
      \end{figure} 
    
      The notebook runs on a docker container with access to 24 CPU cores and 256 Gigabytes of RAM. The docker container
    runs on top of a virtual machine (VM), which is shared among many containers.
    
      Most of the analysis was done running Python code  on the Jupyter notebooks, which allow us to integrate
    data analysis and documentation. Some of the most compute intensive figures were produced on Python running on 
    batch mode instead of inside the notebook, more specifically figures Fig. (\ref{fig:joint-interval-dims}) and
    Fig. (\ref{fig:qr-joint-interval-dims})c, which corresponds to the fractal dimension for the iso-sets for, 
    respectively, joint enstrophy and strainrate, and joint $Q$ and $R$. These calculations required the evaluation of forward and inverse 3D FFTs for 
    each of the $120 \times 120$ geometric sets, i.e. a significant computational effort. To perform the 3D FFTs efficiently, a data-cube must fit in  the RAM of a single compute node.
    
    The present analysis mode shows that, under appropriate circumstances, using Python on Jupyter notebooks within Sciserver is a viable
    option to perform global analysis of large DNS datasets that have been stored in a database such as JHTDB.  


  \section*{Apendix B: Tests of correlation and box-counting on known fractal sets in 3D: Menger Sponge}
    
      In order to validate our techniques and present a known reference for the scaling tools utilized in this work, 
    we present here the results of the correlation-function  and box-counting based analysis for  a known self-similar fractal (the Menger Sponge).
       
      \begin{figure}[ht]
        \begin{minipage}{0.52\linewidth}
          \includegraphics[width=\linewidth]{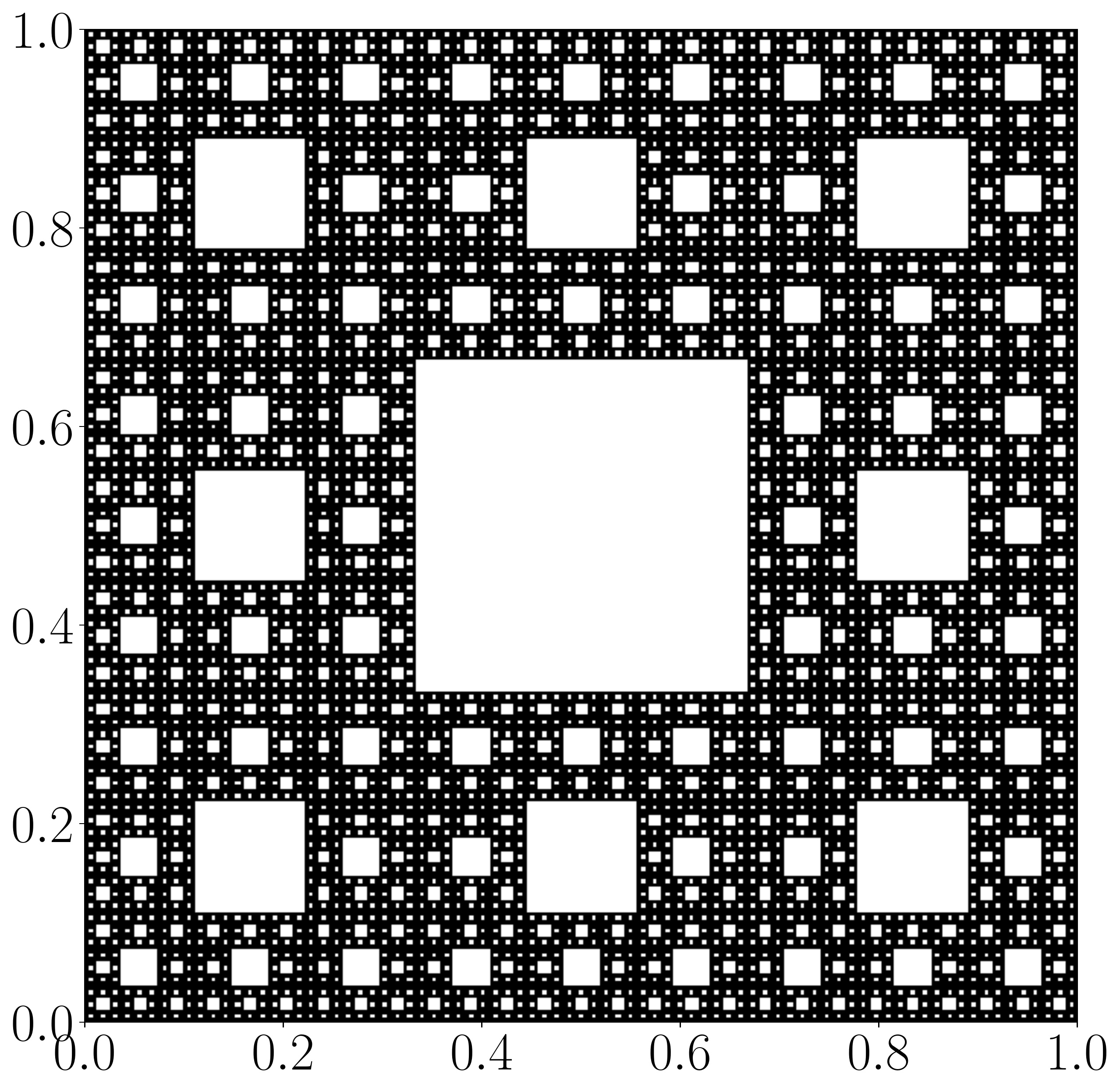}     
          \put (-47,193){\makebox[0.05\linewidth][r]{\Large \color{black}{\Large \bf (a)}}}       
        \end{minipage}
        \begin{minipage}{0.47\linewidth}
          \includegraphics[width=\linewidth]{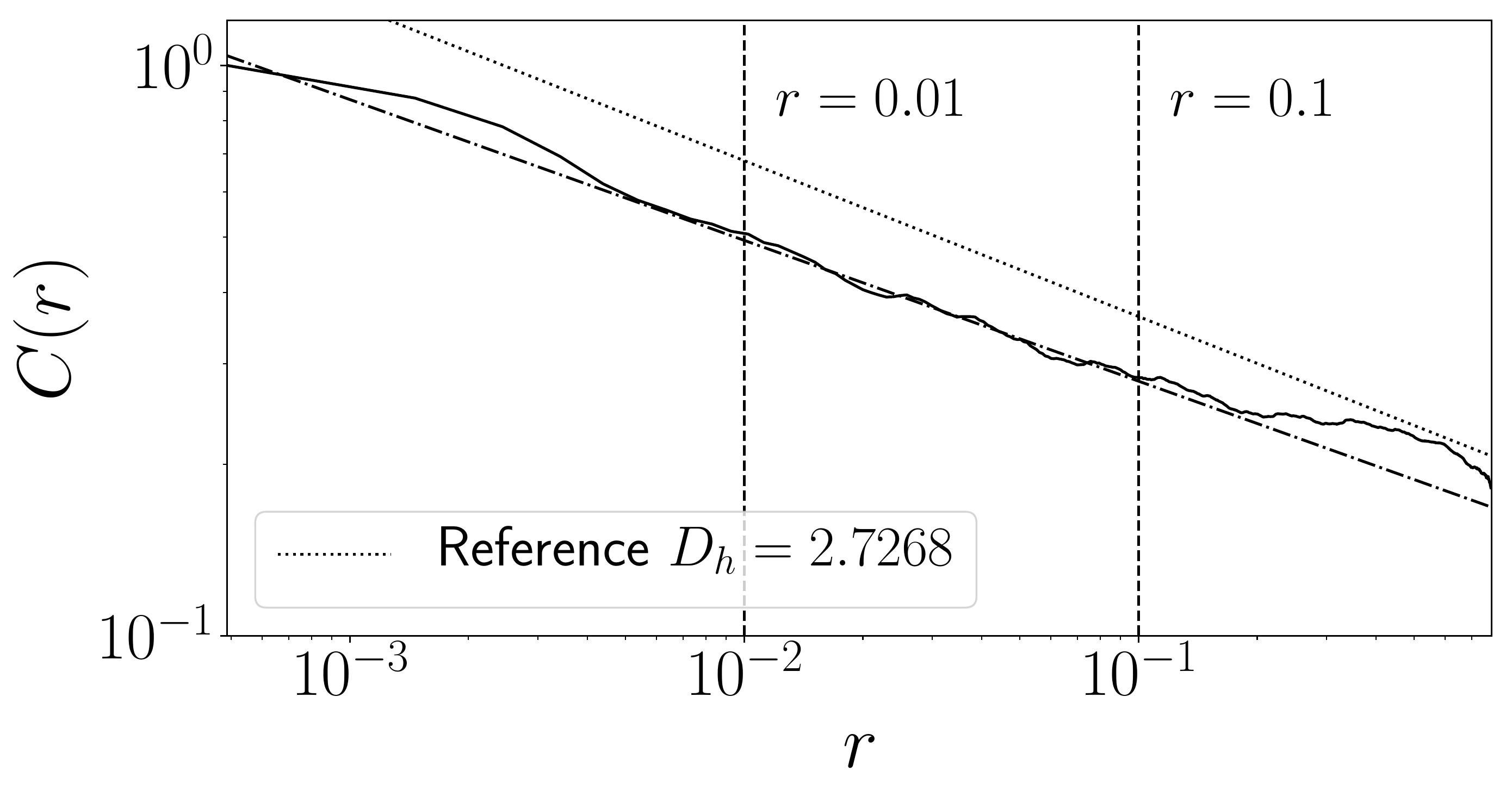}      
          \put (-17,98){\makebox[0.05\linewidth][r]{\Large \color{black}{(b)}}}       
         
          \includegraphics[width=\linewidth]{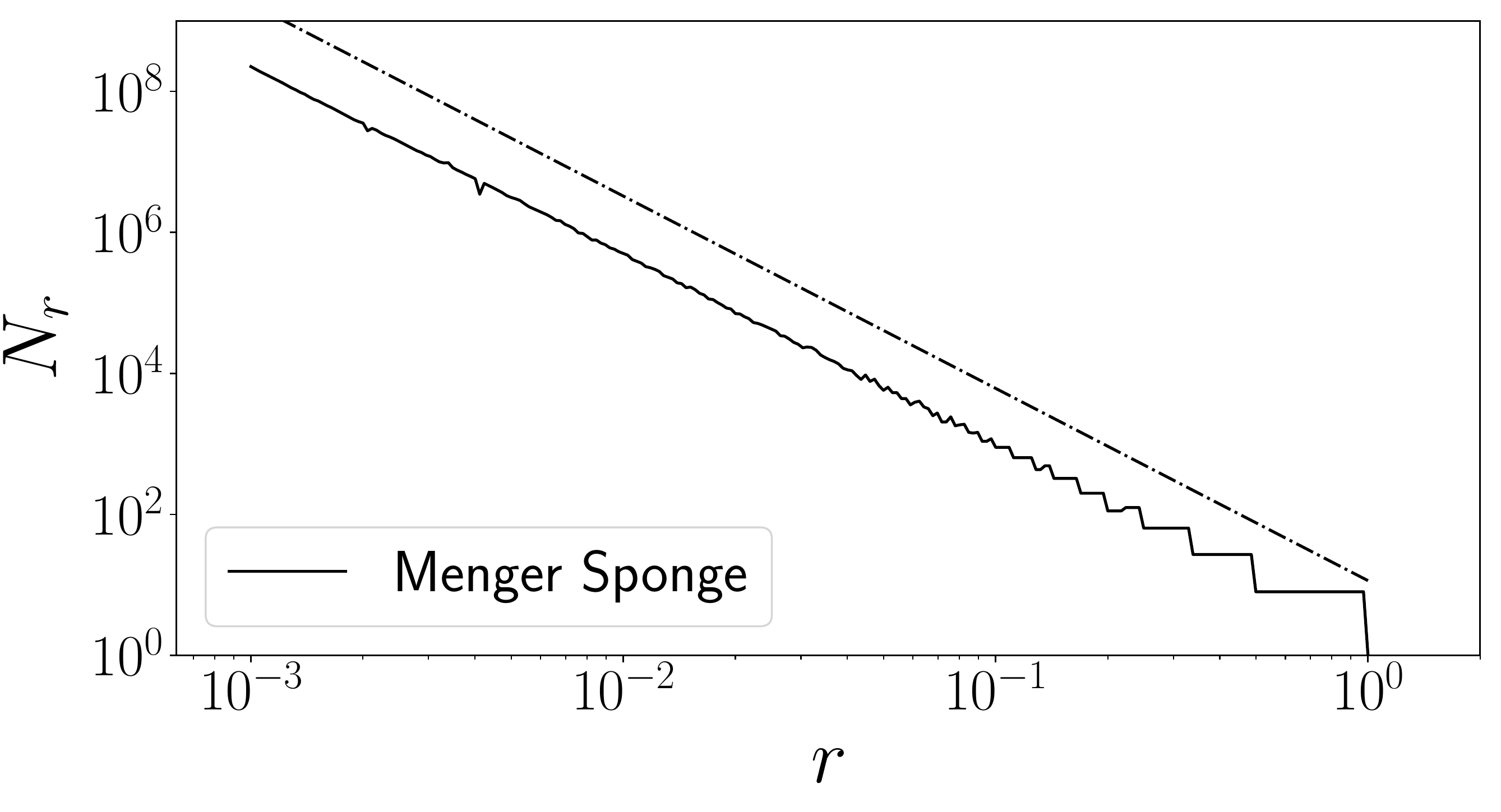}
          \put (-17,100){\makebox[0.05\linewidth][r]{\Large \color{black}{(c)}}}       
        \end{minipage}
       
        \caption{(a): Contour plot of a planar cut through an indicator function marking a 5-level Menger Sponge set; 
                                     (b) Two-point radial correlation function for the Menger Sponge set; (c) Box counting plot. The dotted line is the
                                     slope corresponding to the analytically known fractal dimension of the Menger Sponge ($D_0=D_2 = \log(20)/\log(3)$). }
        \label{fig:menger-sponge}
      \end{figure}     
      
      The Menger Sponge is generated through an iterative process, in which the central 1/3 sized sub-cube  on each of the 6 sides and the core of
    the mother cube are deleted. This process is repeated iteratively for each remaining sub-cube.   
      In our tests, we use  a level 5 Menger sponge, i.e. the 5th iteration of removal as shown in figure
    \ref{fig:menger-sponge}(a). The set indicator function is computed over the same $N^3 = 1024^3$ used for the dataset of 
    this work, in which the removed regions are set to $0$, and the rest is set to $1$ (on elements of scale $1024/3^5 \sim 4$). 
    
      Over this indicator function, we compute the two-point correlation function just as in section \ref{sec:definitions}, 
    which results in figure (\ref{fig:menger-sponge}b). The power-law behavior is affected at large and small scales due to the cubic symmetry of the 
    set being analyzed via spherical bins of distances. Still, there is clearly a  power-law in a central decade in figure \ref{fig:menger-sponge}(b) 
    with a slope consistent with a 
        correlation-function based dimension of $3-\gamma =  {\log 20}/{\log 3}$,   the Haussdorff dimension 
    of the Menger sponge. Analogously, we computed the box-counting graph for 
    the same indicator function. Since the box-counting method is consistent with the artificial fractal set's construction, one 
    obtains a  clearer power law, as seen in figure \ref{fig:menger-sponge}(c).   Again, the slope is consistent with the Haussdorff dimension. These tests verify our   method of computing correlation function and box-counting based scaling exponents.
\end{document}